\documentclass[twocolumn,floatfix,showpacs,prd,aps,eqsecnum,tightenlines,superscriptaddress]{revtex4}
\usepackage{graphicx}
\usepackage{dcolumn}
\usepackage{bm}

\newcommand{\lb}{{\cal L}_\beta}
\newcommand{\dt}{\partial_0}

\newcommand{\bea}{\begin{eqnarray}}
\newcommand{\eea}{\end{eqnarray}}
\newcommand{\beq}{\begin{equation}}
\newcommand{\eeq}{\end{equation}}

\newcommand{\LAZEV}{{\it LazEv}}

\begin{document}


\title{Accurate black hole evolutions by fourth-order numerical relativity}

\author{Y. Zlochower} \affiliation{Department of Physics and Astronomy,
and Center for Gravitational Wave Astronomy,
The University of Texas at Brownsville, Brownsville, Texas 78520}

\author{J. G. Baker} \affiliation{Gravitational Astrophysics Laboratory,  
NASA Goddard Space Flight Center, Greenbelt, Maryland 20771}

\author{M. Campanelli}  \affiliation{Department of Physics and Astronomy,
and Center for Gravitational Wave Astronomy,
The University of Texas at Brownsville, Brownsville, Texas 78520}

\author{C. O. Lousto} \affiliation{Department of Physics and Astronomy,
and Center for Gravitational Wave Astronomy,
The University of Texas at Brownsville, Brownsville, Texas 78520}

\date{\today}

\begin{abstract}
We present techniques for successfully performing numerical relativity
simulations of binary black holes with fourth-order accuracy.  Our
simulations are based on a new coding framework which currently
supports higher order finite differencing for the BSSN formulation
of Einstein's equations, but which is designed to be readily
applicable to a broad class of formulations.  We apply our techniques
to a standard set of numerical relativity test problems, demonstrating
the fourth-order accuracy of the solutions.  Finally we apply our
approach to binary black hole head-on collisions, calculating the
waveforms of gravitational radiation generated and demonstrating
significant improvements in waveform accuracy over second-order
methods with typically achievable numerical resolution.

\end{abstract}

 \maketitle

\section{Introduction}\label{Sec:Intro}

Gravitational wave astronomy will soon provide astrophysicists with a
new tool to observe and analyze some of the most energetic phenomena
in our universe. Collisions of compact objects, such as neutron stars,
stellar-mass black holes, and super-massive black holes, should
produce characteristic gravitational wave signals that are observable
to cosmological distances.  In particular binary black hole systems
are among the most promising sources of gravitational waves for both
the current generation of ground-based detectors, such as 
LIGO~\cite{LIGO}, and for the next generation of space-based detectors,
such as LISA~\cite{LISA,Danzmann:2003tv}. While early ground-based
detectors will need to use
templates from theoretically derived waveforms in order to extract the
weak signal from the much larger noise, space-based detectors will
require accurate models of the gravitational wave sources to extract
the characteristic information (e.g., mass, spin) from the detected
gravitational waves. Thus there is a critical need for accurate
gravitational waveforms from realistic simulations of merging black
holes.

Numerical studies of Einstein's equations play an essential role in
studying highly dynamic systems, such as binary black holes, that have
little or no symmetry. These are not only computationally very
demanding, requiring high-performance massive parallel computers, but
also mathematically and numerically very challenging.  Despite these
difficulties a great deal of progress has been made in the last few
years~\cite{Alcubierre:2004es}, and it is now possible to follow binary 
black hole evolutions during the last moments of the merger
phase~\cite{Brandt00, Alcubierre00b, Alcubierre2003:pre-ISCO-coalescence-times} 
and perhaps even through a complete orbit~\cite{Bruegmann:2003aw}.
The calculation of the gravitational radiation emission from merging 
binary black holes has also been made possible through the use of 
the Lazarus approach, which bridges numerical relativity and perturbative 
techniques to extract approximate gravitational waveforms 
~\cite{Baker:2001nu, Baker:2001sf, 
Baker:2002qf,Baker:2004wv, Campanelli:2004zw}. These calculations
are in good agreement with recent numerical relativity evolutions
~\cite{Alcubierre2003:pre-ISCO-coalescence-times}.

At present, one of the most serious limiting factor in
numerical relativity has been the inability to extract accurate and
reliable results from stable three-dimensional numerical 
simulations of coalescing binary black hole spacetimes. 
Considerable resources have been dedicated to finding numerically 
stable formulations of the Einstein evolution equations 
\cite{Nakamura87,Shibata95, Friedrich96,
Frittelli:1996wr, Baumgarte99, Anderson99, Alcubierre00b,
Kidder01a,Sarbach02b,Shinkai02a, Lindblom:2003ad,Bona:2004yp}. This
effort has led to stable evolutions of single
black hole spacetimes \cite{Alcubierre01a,Yo02a,Alcubierre:2002kk,
Sperhake:2003fc}.
However, there is a growing realization that solving a
well-posed formulation of the Einstein equations with second-order
accurate finite differencing is not sufficient to produce accurate
simulations of binary black hole spacetimes. Unfortunately, black hole
simulations require large domains with high resolution near the
horizons, and the unigrid second-order accurate codes developed in the
past are not sufficient, given the foreseeable constraints on memory
and CPU speed, to produce accurate waveforms from merging black
holes~\cite{Miller:2005qu}.  The Numerical Relativity community is
currently pursuing several different approaches to produce accurate
evolutions. Among them are the use of finite elements, spectral
methods~\cite{Pfeiffer:2002wt,Kidder00a}, adaptive mesh
finite-difference codes~\cite{Imbiriba:2004tp,Fiske:2005fx,Sperhake:2005uf, 
Schnetter-etal-03b,Pretorius:2003wc}, and
higher order finite-difference codes~\cite{Calabrese:2003vx}.
In addition, fourth-order accurate algorithms have recently been designed to evolve
perturbations of nonrotating~\cite{Lousto:2005ip} and rotating black
holes~\cite{Pazos-Avalos:2004rp}. Ideally,
the next generation of finite-difference codes should combine higher
order finite-differencing in combination with mesh refinement
techniques.  However, considerable progress toward highly accurate
evolutions can be achieved using a unigrid higher-order
finite-difference code along with a coordinate system, such as
`Fisheye'~\cite{Baker:2001sf}, that concentrates the gridpoints in the
central region containing the black holes.  In this paper we present a
successful approach to performing fourth-order accurate numerical
relativity simulations.

Our simulations are based on a new numerical relativity evolution code
framework, \LAZEV, to solve the full non-linear Einstein's equations
in 3D.  The \LAZEV\ framework is designed to be modular.  The code
consists of a `Method of Lines' time integrator along with several
Mathematica scripts~\cite{scriptauthors} that convert tensorial
equations into finite-difference algorithms (of arbitrary order) in
C. This modularity allows us to quickly implement new formulations and
gauge choices as needed. We have currently implemented
the ADM formulation as well as several `flavors' of the BSSN
formulation \cite{Nakamura87,Shibata95,Baumgarte99}.  
In Sec.~\ref{sec:lazev} we describe the \LAZEV\
framework in detail.  In Sec.~\ref{sec:formulation} we enumerate many
of the basic techniques utilized in numerical relativity simulations
which we have implemented within \LAZEV, providing the foundation for
detailed testing and application in black hole simulations.

In Sec.~\ref{sec:tests} we test our fourth-order techniques on
standard numerical relativity testbed
problems~\cite{Alcubierre2003:mexico-I} demonstrating fourth-order
accuracy.  Finally, in Sec~\ref{sec:headon} we apply these methods to
study head-on binary black hole collisions. Although head-on collisions
have limited astrophysical relevance, there has been recent interest in
using these spacetimes to develop and test numerical techniques for
evolving more generic binary black hole
spacetimes~\cite{Alcubierre:2004bm,Fiske:2005fx,Sperhake:2005uf}.  In
this paper we demonstrate that our fourth-order techniques can realize
significant improvements in gravitational radiation waveform
accuracies at typically achievable numerical
resolutions.

\section{The \LAZEV\ Evolution Framework}
\label{sec:lazev}

The \LAZEV\ evolution framework is a foundation for rapid
implementation of numerical relativity evolution system routines,
supporting higher order finite differencing techniques with `method of
lines' (MoL) integration~\cite{Gustafsson95}.  Our numerical code uses
the Cactus Computational Toolkit~\cite{cactus_web} for parallelization
and IO. We have constructed the code so that it is compatible with the
initial data and analysis thorns provided with Cactus.  Specific
evolution routines are produced using Mathematica
scripts~\cite{scriptauthors} that convert tensorial equations into finite
difference algorithms in C. 
These scripts provide several types of consistency checks that 
prevent many types of potential coding errors
in setting up the tensorial equations.
The Mathematica scripts support generic centered
spatial finite differencing, with optional upwinded differencing (and
other off-center differencing)
provided by external macros. Currently we use second and fourth-order
spatial finite differencing.

The \LAZEV\ framework applies the MoL technique for solving
first-order-in-time, hyperbolic PDE's, in which the PDE's are converted
into coupled ODE's for all variables at every gridpoint.  This is
achieved by choosing a discrete numerical grid and finite difference
stencils for the spatial derivatives (the finite different stencils
couple the fields at neighboring gridpoints), and then integrating the
time derivatives of the fields at all gridpoints.  This time
integration can be carried out by standard ODE integrators.

The \LAZEV\ MoL integrator provides a generic framework for
integrating hyperbolic PDE's using Runge-Kutta or ICN style time
integrations.  The integrator itself has no knowledge of the system
that is being evolved.  The MoL integrator provides internal timebins
{\tt pre-ministep}, {\tt ministep}, {\tt dissipation}, and {\tt
post-ministep},  which are called during each step of the Runge-Kutta
or ICN integration (there are several {\tt ministeps} in each full
timestep).  The evolution system is chosen by registering routines
with MoL to be evaluated during each of these bins. For example, in
the BSSN system (see Sec.~\ref{sec:evsys}) we register routines to
calculate the time derivative $\partial_t$ of all the BSSN variables in the {\tt
ministep} timebin, a routine to rescale the BSSN conformal metric
$\tilde \gamma_{ij}$ to unit determinant in {\tt post-ministep}, and a
routine to subtract off the numerical trace of the BSSN trace-free, conformal
extrinsic curvature $\tilde A_{ij}$  in {\tt
post-ministep}.  In the ADM system, on the other hand, we register routines
only in the {\tt ministep} bin. Additional evolution systems may also register
routines during these steps. For example, the gauge evolution systems,
which are independent of the main evolution systems, also register
routines in  {\tt ministep} and {\tt post-ministep}.

Currently we have implemented ICN (second-order) and Runge-Kutta
(second, third, and fourth-order). In some cases we use Kreiss-Oliger~\cite{Kreiss73}
 dissipation of the form
\begin{equation}
  \label{eq:ko_diss}
  \partial_t F \to RHS + (-1)^{r/2} \epsilon 
   \sum_i {h_i}^{r+1}{D_{i+}}^{r/2+1}{D_{i-}}^{r/2+1}\,F,
\end{equation}
where $\partial_t F = RHS$ is one of the evolution equations,
$h_i$ is the gridspacing in the $i$th direction, $D_{i+}$ and
$D_{i-}$ are the forward and backwards differencing operators (in the
$i$th direction), and $r$ is the order of the finite differencing used
to evaluate $RHS$.

We use several different styles of finite difference stencils,
depending on the order of accuracy desired and the type of derivative
considered.  Our standard choices for fourth-order accurate evolutions
are:
\begin{eqnarray}
  \partial_x F_{i,j,k} &=& \frac{1}{12 dx}(
    F_{i-2,j,k} - 8 F_{i-1,j,k} \nonumber \\
     &+& 8 F_{i+1, j,k} - F_{i+2,j,k}), \label{eq:dcx}\\
  \partial_{xx} F_{i,j,k} &=& \frac{1}{12 dx^2} (
   - F_{i+2,j,k} + 16 F_{i+1,j,k} \nonumber \\
    &-&30 F_{i,j,k} + 16 F_{i-1,j,k} - F_{i-2,j,k} ),\label{eq:dcxx}\\
  \partial_{xy} F_{i,j,k} &=& \frac{1}{144 dx\,dy} \nonumber \\
    &\,&[
    F_{i-2,j-2,k} - 8 F_{i-1,j-2,k} \nonumber\\
     &+& 8 F_{i+1, j-2,k}- F_{i+2,j-2,k} \nonumber \\
    &-&8(F_{i-2,j-1,k} - 8 F_{i-1,j-1,k} \nonumber \\
    &+& 8 F_{i+1, j-1,k} - F_{i+2,j-1,k}) \nonumber \\
    &+&8(F_{i-2,j+1,k} - 8 F_{i-1,j+1,k} \nonumber \\
    &+& 8 F_{i+1, j+1,k} - F_{i+2,j+1,k}) \nonumber \\
    &-&( F_{i-2,j+2,k} - 8 F_{i-1,j+2,k} \nonumber \\
     &+& 8 F_{i+1, j+2,k} - F_{i+2,j+2,k} \label{eq:dcxy}
) ]
\end{eqnarray}
Note that the mixed $xy$ derivative is obtained by applying the
x-derivative and y-derivatives sequentially (the order is irrelevant).

We do not use standard centered differencing for the advection terms 
(i.e.\ terms of form $\beta^i \partial_i F$). For these terms we use the
following upwinded stencils:
\begin{eqnarray}
  \partial_x F_{i,j,k} &=&\frac{1}{12 dx}\nonumber \\
  &\,& (-F_{i-3,j,k} + 6 F_{i-2,j,k} - 18 F_{i-1,j,k} \nonumber \\
  &+& 10 F_{i,j,k} + 3 F_{i+1,j,k}) \mbox{\ \ \ for
$\beta^x<0$} \label{eq:ddx}\\
  \partial_x F_{i,j,k} &=&\frac{1}{12 dx}\nonumber \\
   &\,& (F_{i+3,j,k} - 6 F_{i+2,j,k} + 18 F_{i+1,j,k} \nonumber \\
   &-& 10 F_{i,j,k} - 3 F_{i-1,j,k}) \mbox{\ \ \ for $\beta^x>0$}
\label{eq:dux}
\end{eqnarray}

In addition we use lower-order centered differencing on the planes
adjacent to the boundaries. We have two choices for how the lower-order 
stencils are constructed. We can choose to use second-order
centered differencing at all points on these planes and for all
directions, or we can use second-order centered differencing only in
the direction perpendicular to the boundary. When we use this latter
choice we construct mixed second derivatives by applying the
fourth-order accurate centered first derivative operator (in a
direction tangent to the boundary) to the second-order centered
derivative operator (perpendicular to the boundary).

Although we plan to implement excision boundary conditions soon,
we currently use the puncture approach
(see Ref.~\cite{Alcubierre:2004bm} for a comparison of methods),
along with singularity avoiding 
slicings, to evolve black hole space-times. 
When using punctures, we found that we needed to either use second-order
stencils (but still fourth-order Runge-Kutta time integration) in
regions inside the apparent horizons, or use a second-order upwinded
stencil for the advection terms over the entire computational
domain. The former method produces significantly better waveforms. See
Sec.~\ref{sec:4thRes} for further details.
 
\section{Formulation}
\label{sec:formulation}
Within our \LAZEV\ framework we have implemented support for several
standard options in formulating numerical relativity modeling problems.
Such a formulation includes selection of an Einstein evolution system
and a choice of gauge for the evolving spacetime.  In this section we 
discuss the formulation options presently available within our \LAZEV\ 
framework.  Of these we have specific realizations which are appropriate 
for the fourth-order applications discussed in the subsequent sections.

\subsection{Evolution Systems}
\label{sec:evsys}

Many alternative formulations of Einstein's evolution equations have
been considered so far (see Ref.~\cite{Alcubierre:2000ke} and
references therein). Within the most popular Cauchy or $3+1$ approach,
one can currently choose among the first order symmetric hyperbolic
formulations of the evolution equations which are mathematically very
attractive~\cite{Kidder01a,Sarbach02b,Lindblom:2003ad,Bona:2004yp},
and various flavors of spatially second-order hyperbolic formulations which are
numerically more tractable~\cite{Nagy:2004td,Gundlach:2004jp}. While
the former have the advantage of possessing a well-posed continuum
limit in the presence of maximally dissipative boundary conditions, it
is not yet clear how one can handle dozens of free
parameters~\cite{Kidder01a,Sarbach02b} and dynamical gauge choices for
the lapse and shift. On the other hand, some second-order hyperbolic
formulations have proven to be empirically much more robust than
others~\cite{Shibata95,Baumgarte99,Pretorius:2004jg, Kreiss:2002}.  In
this paper we will focus on the Baumgarte-Shapiro-Shibata-Nakamura
(BSSN)~\cite{Nakamura87,Shibata95,Baumgarte99} system, a strongly
hyperbolic system~\cite{Sarbach02a,Beyer:2004sv} that has been shown
to have some attractive stability
properties~\cite{Alcubierre99e,Sarbach02a,Beyer:2004sv}.

The BSSN system is an extension of
the standard ADM~\cite{Arnowitt62} system with better numerical
properties than the original system. In the ADM system the Einstein
Equations are split into six evolution equations for the metric (along
with six auxiliary evolution equations for the extrinsic curvature)
\begin{eqnarray}
\dt \gamma_{ij} &=& - 2 \alpha K_{ij},
\label{dgdt} \\
\dt K_{ij} &=& - D_i D_j \alpha,
\nonumber \\
&&  + \alpha (R_{ij} + K K_{ij} - 2 K_{ik} {K^k}{}_j),
\label{dKdt}
\end{eqnarray}
and four constraints equations
\begin{eqnarray}
{\cal H} & \equiv & R + K^2 - K_{ij} K^{ij} = 0,
\label{Hconstraint}\\
{\cal M}^i & \equiv & D_j (K^{ij} - \gamma^{ij} K) = 0.
\label{Dconstraint}
\end{eqnarray}
Here $\dt $ is the operator $\partial_t - \lb$, $\lb$ is the Lie
derivative with respect to the shift vector $\beta^i$, $D_i$ is the
covariant derivative associated with the 3-metric $\gamma_{ij}$,
$R_{ij}$ is the three-dimensional Ricci tensor, $R$ the Ricci scalar,
and $K$ is the trace of $K_{ij}$.

The BSSN system of equations is obtained from
the standard ADM equations by the following
substitutions,
\begin{eqnarray}
  \gamma_{ij} &\to& e^{4 \phi} \tilde \gamma_{ij}, \\
   K_{ij} &\to& e^{4 \phi} \left( \tilde A_{ij} +
    \frac{1}{3} \tilde \gamma_{ij} K\right),
\end{eqnarray}
where $\tilde \gamma = \det \tilde \gamma_{ij} = 1$ and
 $\tilde A^{i}_{\,i} = 0$.
Three additional variables
\begin{equation}
  \tilde \Gamma^i = -\partial_j \tilde \gamma^{i j}
\end{equation}
are also introduced. The BSSN variables ($\tilde \gamma_{i j}$, $\tilde
A_{ij}$, $K$, $\phi$, and $\tilde \Gamma^i$) obey the following
evolution equations~\cite{Alcubierre02a}
\begin{eqnarray}
   \label{eq:gt_evol}
   \dt \tilde \gamma_{ij} &=& -2 \alpha \tilde A_{ij}, \\
   \dt \phi &=& -\frac{1}{6} \alpha K, \\
   \dt \tilde A_{ij} &=& e^{-4 \phi}\left(-D_i D_j \alpha +
       \alpha R_{ij}\right)^{TF} +\nonumber \\
        &\,&\alpha \left(K \tilde A_{ij} -
        2 \tilde A_{ik} \tilde A^{k}_{\,j}\right), \\
   \dt K &=& - D^i D_i \alpha + \alpha \left(\tilde A_{ij}\tilde
         A^{ij} +\frac{1}{3}K^2\right),\\
  \partial_t \tilde \Gamma^i &=& \tilde \gamma^{jk} \partial_j
\partial_k \beta^i + \frac{1}{3} \tilde \gamma^{ij} \partial_j
\partial_k \beta^k + \beta^j \partial_j \tilde \Gamma^i - \nonumber \\
  &\,&\tilde
\Gamma^j \partial_j \beta^i + 
 \frac{2}{3}\tilde \Gamma^i \partial_j
\beta^j - 2 \tilde A^{i j}\partial_j \alpha + \nonumber \\
 &\,& 2 \alpha \left(\tilde
{\Gamma^i}_{jk} \tilde A^{jk} + 6 \tilde A^{ij}\partial_j \phi -
\frac{2}{3} \tilde \gamma^{ij} \partial_j K\right), \hspace{10mm}  \label{eq:dtGamma}
\end{eqnarray}
where $TF$ indicates that only the trace-free part of the tensor is
used and $R_{ij} = \tilde R_{ij} + {R^\phi}_{ij}$ is given by
\begin{eqnarray}
  {R^\phi}_{ij} &=& - 2 \tilde D_i \tilde D_j \phi - 2 \tilde
\gamma_{ij} \tilde D^k \tilde D_k \phi + 4 \tilde D_i \phi \tilde D_j \phi
- \nonumber \\
  &\,&4\tilde \gamma_{i j} \tilde D^k \phi \tilde D_k \phi, \\
 \label{eq:RT}
 \tilde R_{ij} &=& -\frac{1}{2} \tilde \gamma^{lm}\partial_l
\partial_m \tilde \gamma_{ij} + \tilde \gamma_{k(i}\partial_{j)}
\tilde \Gamma^k + \tilde \Gamma^k \tilde \Gamma_{(ij)k} + \nonumber \\
 &\,&\tilde
\gamma^{lm}\left(2 \tilde {\Gamma^k}_{l(i}\tilde \Gamma_{j)km} +
\tilde {\Gamma^k}_{im}\tilde \Gamma_{klj}\right),
\end{eqnarray}
and $\tilde D_i$ is the covariant derivative with respect to $\tilde
\gamma_{ij}$.
$\tilde \Gamma^i$ is replaced by $-\partial_j \tilde
\gamma^{ij}$ in Eq's~(\ref{eq:gt_evol}) - (\ref{eq:RT})
wherever it is not differentiated. Note that Eq.~(\ref{eq:dtGamma})
gives the $\partial_t$ derivative of $\tilde \Gamma^i$ rather than the
$\partial_0$ derivative. The Lie derivatives of the non-tensorial
quantities ($\phi$, $\tilde \gamma_{ij}$, and $\tilde A_{ij}$) are given by
\begin{eqnarray}
  \lb \phi &=& \beta^k \partial_k \phi + \frac{1}{6} \partial_k \beta^k, \\
  \lb \tilde \gamma_{ij} &=& \beta^k\partial_k \tilde \gamma_{ij} +
   \tilde \gamma_{ik} \partial_j \beta^k + \nonumber \\
     &\,&\tilde \gamma_{jk} \partial_i \beta^k -
    \frac{2}{3} \tilde \gamma_{ij} \partial_k \beta^k, \\
  \lb \tilde A_{ij} &=& \beta^k\partial_k \tilde A_{ij} +
   \tilde A_{ik} \partial_j \beta^k + \nonumber \\
     &\,&\tilde A_{jk} \partial_i \beta^k -
    \frac{2}{3} \tilde A_{ij} \partial_k \beta^k.
\end{eqnarray}

In addition to the evolution equations, the BSSN variables must also
obey the following seven differential constraint equations
\begin{eqnarray}
  \label{eq:ham_const}
  {\cal H} &=& R - \tilde A_{ij} \tilde A^{ij} + \frac{2}{3}K^2 = 0, \\
  \label{eq:mom_const}
  e^{4 \phi} \,{\cal M}^i &=& \partial_j \tilde A^{ij} + \tilde {\Gamma ^i}_{jk} \tilde
A^{jk} + 6 \tilde A^{ij} \partial_j \phi - \nonumber\\
  &\,&\frac{2}{3} \tilde
\gamma^{ij} \partial_j K = 0, \\
  \label{eq:bssn_const}
  {\cal G}^i &=& \tilde \Gamma^i + \partial_j \tilde \gamma^{i j} = 0,
\end{eqnarray}
and the following two algebraic constraint equations
\begin{eqnarray}
\tilde \gamma &=&1, \label{eq:detconst}\\
\tilde A^{i}_{\,i}&=&0. \label{eq:trconst}
\end{eqnarray}
We monitor the differential constraints but do not enforce them.
However,
we enforce the algebraic constraints by rescaling the evolved $\tilde
\gamma_{ij}$ and subtracting off the trace of the evolved $\tilde
A_{ij}$ at every timestep.

We can evaluate $(R_{ij})^{TF} = R_{ij} - 1/3 \gamma_{ij} R$ in two
ways.  We could either calculate $R_{ij}$ and find its trace $R$
numerically and subtract off the numerical trace, or we can use
the Hamiltonian constraint Eq.~(\ref{eq:ham_const}) to calculate $R$ and
subtract that from $R_{ij}$. Our tests showed very little difference
between the two approaches even when the Hamiltonian constraint
violations were relatively large. Unless otherwise specified we use the
latter method.

\subsection{Gauge Choices}
\label{sec:gauge}
We have implemented three basic types of lapse conditions: (i) maximal
slicing, (ii) ``Bona-Mass{\'o}'' type lapses~\cite{Bona94b,Alcubierre01a,
Balakrishna96a,Alcubierre02a}, (ii) and
densitized lapses.
The maximal slicing condition has the form
\begin{equation}
  \label{eq:maximal}
  \Delta \alpha = \beta^i \partial_i K + \alpha K_{ij} K^{ij},
\end{equation}
and is implemented using a modified version of the Cactus `Maximal'
thorn~\cite{cactus_web} in
combination with a fourth-order accurate version of
Bernd Br{\"u}gmann's `BAM\_Elliptic'~\cite{Brandt97b,cactus_web}
thorn. This fourth-order accurate version of `BAM\_Elliptic' was
developed at UTB by Mark Hannam.
The ``K-driver'' lapses have the form
\begin{equation}
  \label{eq:kdriver}
  \partial_t \alpha = - \alpha^2 f(\alpha) A,
\end{equation}
where
\begin{equation}
  \label{eq:kdriver_A}
  \partial_t A = \partial_t K - \xi A,
\end{equation}
and $\xi$ is some specified function (usually zero).
We have also implemented modifications to this general form by replacing
$\partial_t \alpha$ with $\partial_0 \alpha$, adding
$D_i \beta^i$ to the right-hand-side of Eq.~(\ref{eq:kdriver}),
and multiplying $A$ by factors of $\psi^n$ in  Eq.~(\ref{eq:kdriver}),
where $\psi$ is the conformal
factor of the puncture data.
Equation~(\ref{eq:kdriver}) includes harmonic slicing ($f(\alpha) = 1$),
{\tt 1+log} slicing ($f(\alpha) = 2/\alpha$), and `shock-avoiding' slicing
($f(\alpha) = \frac{8}{3} (\alpha(3 - \alpha))^{-1}$)~\cite{Alcubierre02b}.

We have implemented `Gamma-driver' shift conditions~\cite{Alcubierre02a}
as well as static corotating shifts. The two
versions of the `Gamma-driver' shifts that we implemented are
\begin{eqnarray}
   \partial_t \beta^i &=& F\, B^i -\zeta \,\beta^i,\nonumber \\
   \partial_t B^i &=& \partial_t \tilde \Gamma^i - \eta B^i,
\end{eqnarray}
and
\begin{eqnarray}
   \label{eq:gamma_driver_alt}
   \partial_t \beta^i &=& B^i -\zeta \,\beta^i,\nonumber \\
   \partial_t B^i &=& F\,\partial_t \tilde \Gamma^i - \eta B^i,
\end{eqnarray}
where $\eta$ and $\zeta$ are some prescribed functions over space, and $F$
is a function of $\alpha$ and $\psi$.
Unless otherwise specified, we take $\zeta = 0$.

\subsection{Boundaries}

Thus far we have implemented fairly simple boundary conditions. For
fourth-order upwinded stencils we alternatively use centered finite 
differencing or second-order upwinding at the
second point from the boundary, and second-order centered differencing
at the first point from the boundary. The boundary points are filled
using radiative boundary conditions~\cite{Alcubierre02a}. However,
when the shift is zero,
or sufficiently small, we evolve $\tilde \gamma_{ij}$ on the boundary
using the zero-shift form of Eq.~(\ref{eq:gt_evol}).
We have observed that the boundary algorithm has very little effect on
the convergence of the $\ell=2$ component of the waveforms in Sec.~\ref{sec:headon}.

These boundary conditions are not known to be well posed and do lead to
incoming constraint violating modes. Our future work will involve
imposing constraint-preserving boundary 
conditions~\cite{Gundlach:2004jp, Frittelli:2004, Calabrese01a}.

\section {Apples With Apples Tests}
\label{sec:tests}

We apply the \LAZEV\ framework to a standard set of numerical relativity
tests known as `Apples with Apples' tests~\cite{Alcubierre2003:mexico-I}.
These tests consists of evolving
spacetimes with $R\times T^3$ topology with the advantage that
there are no boundaries. The results from these testbeds indicate
that \LAZEV\ is stable and fourth-order convergent with our choices of
fourth-order stencils.

\subsection{Gauge wave test}
For this test we evolve the metric 
\begin{equation}
  ds^2 = - H\,dt^2 + H\, dx^2 + dy^2 + dz^2,
\end{equation}
where
$$
 H = 1 + A\,\sin\left(\frac{2 \pi (x-t)}{d}\right),
$$
and $d$ is the wavelength.
The ADM variables for this metric have the form
\begin{eqnarray}
  \gamma_{xx} &=& 1 + A\,\sin\left(\frac{2 \pi (x-t)}{d}\right), \\
  \gamma_{yy} &=&\gamma_{zz} = 1, \\
   K_{xx} &=& \frac{A \pi}{d} \frac{\cos\left(\frac{2
\pi(x-t)}{d}\right)}{\sqrt{1+A\, \sin\left(\frac{2
\pi(x-t)}{d}\right)}}, \\
  \alpha &=& \sqrt{1 + A\,\sin\left(\frac{2\pi(x-t)}{d}\right)},
\end{eqnarray}
where all remaining ADM variables (including $\beta^i$) are zero.
Note that $\alpha$ obeys the harmonic slicing condition 
\begin{equation}
  \partial_t \alpha = - \alpha^2 K.
\end{equation}
In our simulations $\alpha$ is `live' and we evolve it using the
harmonic slicing condition. A two-dimensional test is obtained by the
coordinate transformation
\begin{eqnarray*}
  x \to \frac{1}{\sqrt{2}}(x+y),\\
  y \to \frac{1}{\sqrt{2}}(y-x),
\end{eqnarray*}
producing non-trivial $\gamma_{xx}$, $\gamma_{yy}$, $K_{xx}$, and
$K_{yy}$.

In order to obtain stable runs we needed to include 
dissipation of the form given in Eq.~(\ref{eq:ko_diss}), with dissipation
coefficient $\epsilon=0.125$.

These one and two-dimensional problems were evolved using a
three-dimensional grid with periodic boundary conditions in all
directions. We chose a wavelength ($d$) of 1 and constructed the
grid so that it contained  $6+1/h$ points
(6 points for ghost-zones) in  the non-trivial directions and 
9 points in the trivial directions
(the stencil requires 3 ghost-zones), where $h$ is the gridsize and
$1/h$ is an integer.

\begin{figure}
\begin{center}
\includegraphics[width=3.2in]{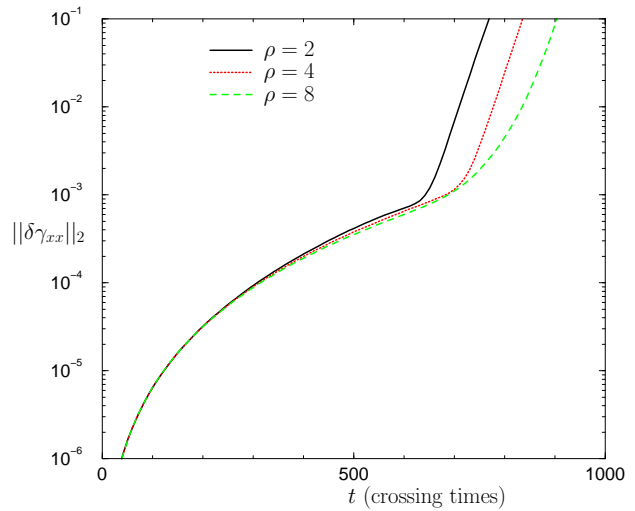}
\end{center}
\caption{The $L_2$ norm of $\delta \gamma_{xx} = \gamma_{xx} -
\gamma_{xx}^{analytic}$, rescaled by
$\rho^4/16$, 
  for the one-dimensional `Gauge Wave' test with $A=0.01$.
  Note the near perfect overlap
  for 630 crossing times and that the larger $\rho$ runs are 
convergent longer. The runs are acceptably accurate when the
(non-rescaled) norm of the error is smaller than $10^{-4}$ (i.e.\ $1\%$ of $A$).
}
\label{fig:gw_gxx_conv}
\end{figure}
Our first test is a weak gauge wave with amplitude $A=0.01$ and
resolutions $h  = 0.02/\rho$, where $\rho=$ $2$, $4$, $8$.
Figure~\ref{fig:gw_gxx_conv} shows the $L_2$ norm of the error in
$\gamma_{xx}$ versus time (rescaled by a factor of $\rho^4/16$).
Note that fourth-order convergence (as evident by the
overlap of the rescaled curves) implies that the error for the $\rho=8$ 
case is 256 times smaller than the $\rho=2$ case. This relationship 
breaks down near $630$ crossing
times, though the higher resolution runs remain 
in the convergence regime longer. Using the criterion that the numerical
solution is sufficiently accurate if the norm of the error is smaller
than
$1\%$ of the amplitude $A$, we find that the fourth-order code produces
accurate results to $t=308$, $t=720$, and $t=865$ for the $\rho=2$,
$\rho=4$, and $\rho=8$ runs respectively. 
\begin{figure}
\begin{center}
\includegraphics[width=3.2in]{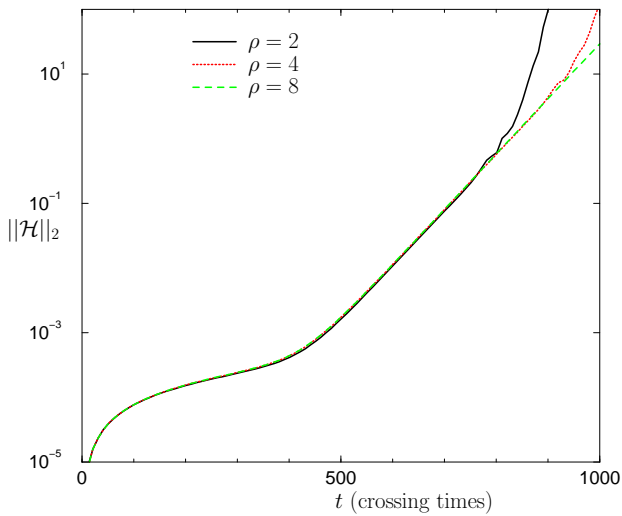}
\end{center}
\caption{The $L_2$ norm of ${\cal H}$, rescaled by $\rho^4/16$, for the
one-dimensional `Gauge Wave' test with $A=0.01$.
  Note the near perfect overlap
  for 800 crossing times and that the larger $\rho$ runs are 
convergent longer. The runs are acceptably accurate when the
(non-rescaled) norm of the Hamiltonian constraint is smaller than
$10^{-4}$ (i.e.\ $1\%$ of $A$) }
\label{fig:gw_HC_conv}
\end{figure}
Figure~\ref{fig:gw_HC_conv} shows the $L_2$ norm of
${\cal H}$ (rescaled by a factor of $\rho^4/16$).
Again, the higher resolution runs remain in the convergence regime
longer. Using the criterion that the numerical
solution is sufficiently accurate if the norm of the Hamiltonian
constraint  is smaller than
$1\%$ of the amplitude $A$, we find that the fourth-order code produces
accurate results to $t=130$, $t=495$, and $t=638$ for the $\rho=2$,
$\rho=4$, and $\rho=8$ runs respectively. The Hamiltonian constrain
violation is a stricter measure of the quality of the results than the
error in $\gamma_{xx}$ because the Hamiltonian constraint
involves second derivatives of the metric (which are harder to model).

\begin{figure}
\begin{center}
\includegraphics[width=3.2in]{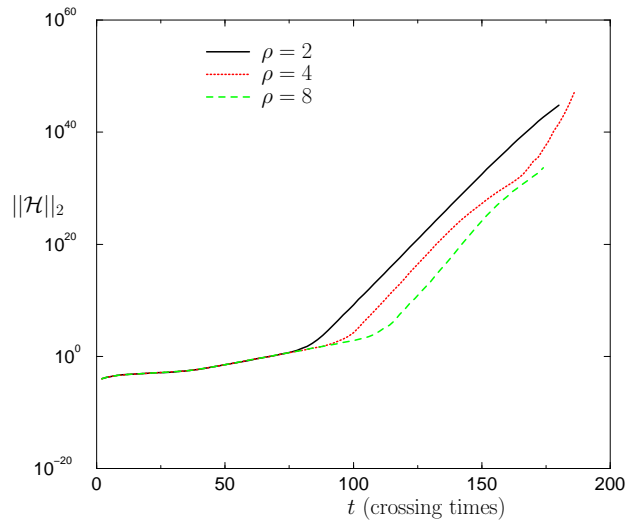}
\end{center}
\caption{The $L_2$ norm of ${\cal H}$, rescaled by $\rho^4/16$, for the
one-dimensional `Gauge Wave' test with $A=0.1$.
  Note the near perfect overlap
  for 80 crossing times and that the larger $\rho$ runs are 
convergent longer. The runs are acceptably accurate when the
(non-rescaled) norm of the Hamiltonian constraint is smaller than $10^{-3}$
(i.e.\ $1\%$ of $A$). }
\label{fig:gw_HC_conv_A01}
\end{figure}

Our next test is a stronger gauge wave.
Figure~\ref{fig:gw_HC_conv_A01} shows the $L_2$ norm of
${\cal H}$ versus time (rescaled by a factor of $\rho^4/16$) for
the gauge-wave test with amplitude $A=0.1$ and
resolutions $h  = 0.02/\rho$, where $\rho=$ $2$, $4$, $8$.
Note that in this higher amplitude case the runs remain convergent for
80 crossing times (approximately one-tenth as long as the $A=0.01$
runs). Using the criterion that the numerical
solution is sufficiently accurate if the norm of the Hamiltonian
constraint  is smaller than
$1\%$ of the amplitude $A$, we find that the fourth-order code produces
accurate results to $t=19.5$, $t=46$, and $t=60$ for the $\rho=2$,
$\rho=4$, and $\rho=8$ runs respectively. Hence these $A=0.1$ runs
are accurate for roughly one-tenth the time that the $A=0.01$ are
accurate.

\begin{figure}
\begin{center}
\includegraphics[width=3.2in]{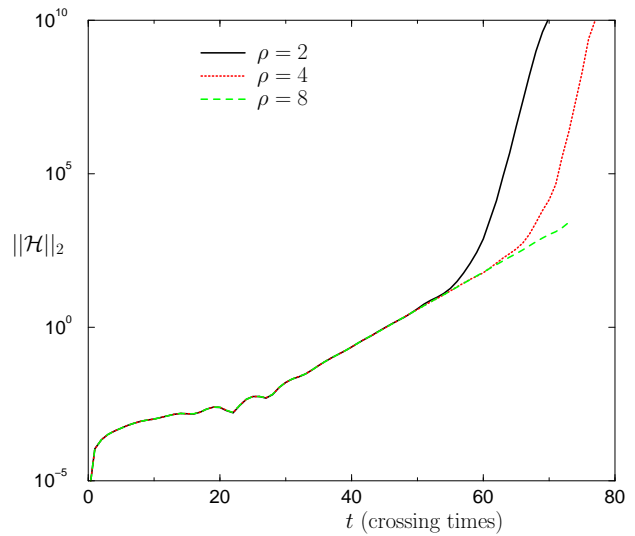}
\end{center}
\caption{The $L_2$ norm of ${\cal H}$, rescaled by $\rho^4/16$, for the
two-dimensional `Gauge Wave' test with $A=0.1$.
  Note the near perfect overlap
  for 55 crossing times and that the larger $\rho$ runs are 
convergent longer. The runs are acceptably accurate when the
(non-rescaled) norm of the Hamiltonian constraint is smaller than $10^{-3}$
(i.e.\ $1\%$ of $A$).  }
\label{fig:gw_HC_conv_A01_2d}
\end{figure}

Though we have applied our fully three-dimensional code to the last
two test problems, the dynamics in these cases are non-trivial only in 
one (numerical grid) direction.  Our last gauge wave test is non-trivial in
two grid directions. 
Figure~\ref{fig:gw_HC_conv_A01_2d} shows the $L_2$ norm of
${\cal H}$ versus time (rescaled by a factor of $\rho^4/16$) for
the two-dimensional gauge-wave test with amplitude $A=0.1$ and
resolutions $h  = 0.02/\rho$, where $\rho=$ $2$, $4$, $8$.
These two-dimensional runs are less stable than the corresponding
one-dimensional run, crashing before 100 crossing times.  Convergence 
is, nonetheless, maintained for about 55 crossing times, longer for 
the higher resolution runs. 
Using the criterion that the numerical
solution is sufficiently accurate if the norm of the Hamiltonian
constraint  is smaller than
$1\%$ of the amplitude $A$, we find that the fourth-order code produces
accurate results to $t=10$, $t=30$, and $t=40$ for the $\rho=2$,
$\rho=4$, and $\rho=8$ runs respectively. Note that these two-dimensional
$A=0.1$ runs are accurate roughly two-thirds as long as the corresponding 
one-dimensional $A=0.1$ runs.

\subsection{Gowdy wave test}

In this section we present results for the `Polarized Gowdy Wave' test.
The `Polarized Gowdy Wave' metric is given by
\begin{equation}
 ds^2 = t^{-1/2} e^{\lambda/2}(-dt^2 + dz^2) + t (e^P dx^2 + e^{-P}
dy^2),
\end{equation}
where
\begin{eqnarray}
  P &=& J_0(2 \pi t) \cos(2 \pi z),\\
  \lambda &=& -2 \pi t J_0(2 \pi t) J_1(2 \pi t) \cos^2(2 \pi z) +
\nonumber \\
&\,& 2 \pi^2 z^2\left[J_0^2(2 \pi t) + J_1^2(2 \pi t)\right] - \nonumber \\
&\,& \frac{1}{2}((2 \pi)^2[J_0^2(2 \pi) + J_1^2(2 \pi)] -\nonumber \\
&\,& 2 \pi J_0(2 \pi) J_1(2\pi)).
\end{eqnarray}
We use this metric to obtain initial data and evolve backwards in time
using the harmonic slicing condition. The above metric only varies as a
function of $z$ and $t$ but we can obtain a two-dimensional problem by introducing
new coordinates $x = \tilde x$, $y=\tilde y$, $z=\tilde z + \tilde y$.

Here again we use a full three-dimensional grid to solve one and two-dimensional
problems. We use periodic boundary conditions in all directions, and our
grid consists of $1/h$ points (plus 6 for ghost-zones) in the
non-trivial directions, and 9 points in the trivial directions.
Additionally, we needed to add dissipation of the form given in
Eq.~(\ref{eq:ko_diss}), with dissipation coefficient
$\epsilon=0.000625$, to stabilize the runs.

\begin{figure}
\begin{center}
\includegraphics[width=3.2in]{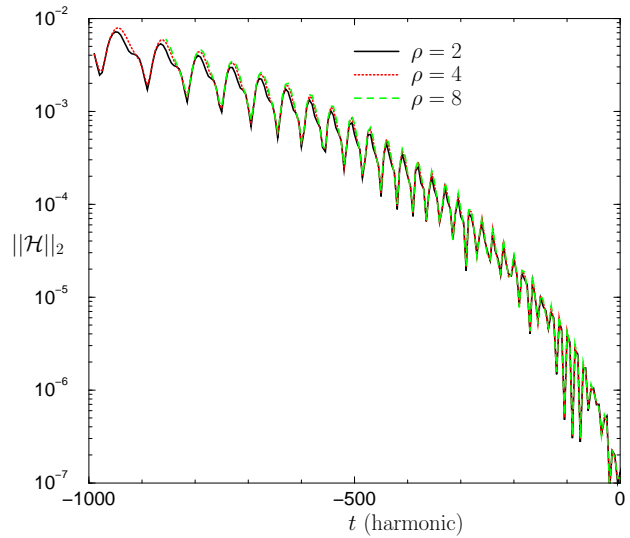}
\end{center}
\caption{The $L_2$ norm of ${\cal H}$, rescaled by $\rho^4/16$, for the
one-dimensional `Gowdy Wave' test.
  Note the good agreement between the curves for 1000 
  crossing times and that the evolution is backwards in time}
\label{fig:gowdy_HC_conv}
\end{figure}

\begin{figure}
\begin{center}
\includegraphics[width=3.2in]{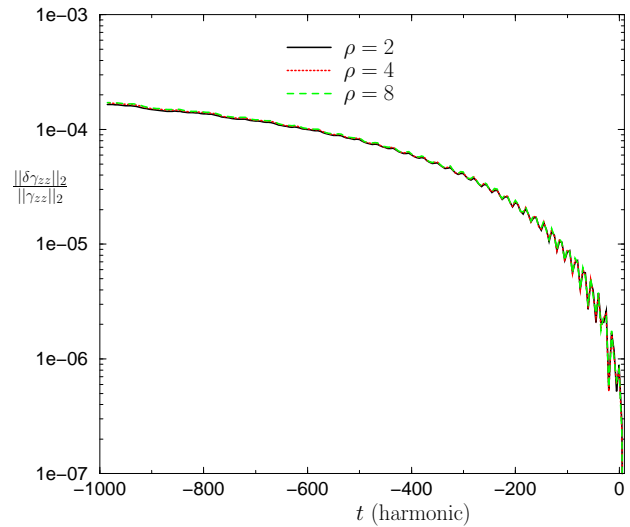}
\end{center}
\caption{The $L_2$ norm of the error in $\gamma_{zz}$, rescaled by the
$L_2$ norm of $\gamma_{zz}$ and by
$\rho^4/16$, for the
one-dimensional `Gowdy Wave' test.
  Note the good agreement between the curves for 1000 
  crossing times and that the relative error in $\gamma_{zz}$ is less
than $2\cdot 10^{-4}$ at
1000 crossing times.
The evolution is backwards in time.}
\label{fig:gowdy_gzz_conv}
\end{figure}

Figures~\ref{fig:gowdy_HC_conv} and \ref{fig:gowdy_gzz_conv} show the
fourth-order convergence of the Hamiltonian constraint and the $\gamma_{zz}$
component of the metric of the one-dimensional Gowdy Wave test for 1000
crossing times. Note that these convergence plots imply that the errors
in $\gamma_{zz}$ and ${\cal H}$ in the $\rho=8$ run are 256 times
smaller than these errors in the $\rho=2$ run. Unlike the previous
gauge-wave test, here the amplitude of the metric functions are
damped. So our criterion for an acceptable value of the Hamiltonian
constrain violation is time dependent.
In this case we will use the criterion that the error norm, divided by
the norm of $\gamma_{zz}$, must be smaller than $1\%$.
 At the end of the run the $L_2$ of $\gamma_{zz}$ (the function,
not the error)  is of order 1. Both the 
Hamiltonian constraint and the error in $\gamma_{zz}$ are smaller than
this quality criterion for all resolutions.

\begin{figure}
\begin{center}
\includegraphics[width=3.2in]{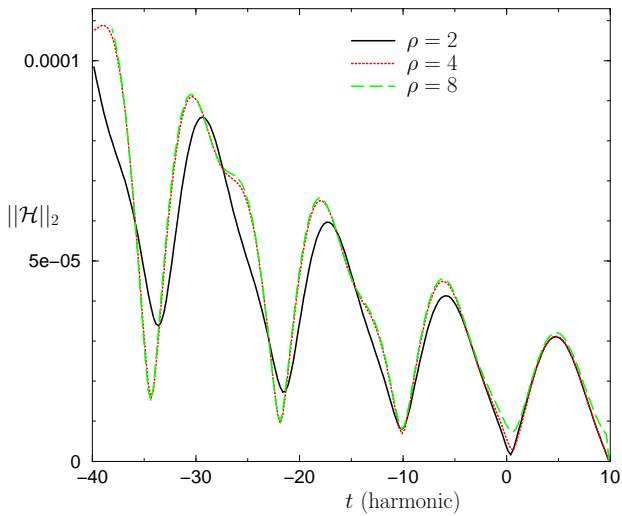}
\end{center}
\caption{The $L_2$ norm of ${\cal H}$, rescaled by $\rho^4/16$, for the
two-dimensional `Gowdy Wave' test.
  The test was limited to 50 crossing times due to limited resources.
  The early time lack of convergence is due to low amplitude high
frequency noise in the $\rho=8$ results. Although the two higher
resolution curves do
not lie on top of $\rho=2$ curve, the overlap of the two high resolution
curves indicates that the Hamiltonian constraint is converging to
fourth-order. Note that the evolution is backwards in time.}
\label{fig:gowdy_HC_conv_2d}
\end{figure}

Figure~\ref{fig:gowdy_HC_conv_2d} shows the fourth-order convergence of
the Hamiltonian constraint for the two-dimensional Gowdy Wave test.
The early time lack of convergence is due to roundoff effects,
 and, although the $\rho=2$ curve does not lie on the $\rho=4$ and
$\rho=8$ curves, the overlap of the $\rho=4$ and $\rho=8$ curves confirm
that the code is fourth-order convergent with sufficient resolution.

\begin{figure}
\begin{center}
\includegraphics[width=3.2in]{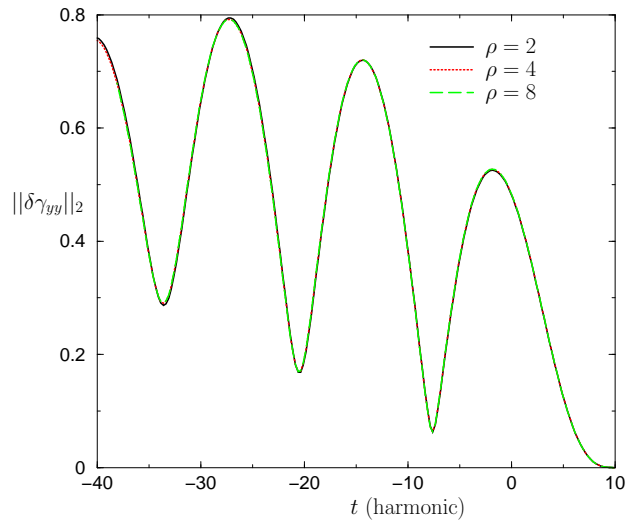}
\end{center}
\caption{The $L_2$ norm of the error in $\gamma_{yy}$, rescaled by
$\rho^4/16$, for the
two-dimensional `Gowdy Wave' test.
  The test was limited to 50 crossing times due to limited resources.
The excellent overlap of all three curves indicates that the code is
converging to fourth-order. Note that the evolution is backwards in time.}
\label{fig:gowdy_gyy_conv_2d}
\end{figure}

Figure~\ref{fig:gowdy_gyy_conv_2d} shows the fourth-order convergence of
$\gamma_{yy}$ for the two-dimensional Gowdy Wave test. Unlike ${\cal H}$,
$\gamma_{yy}$ clearly demonstrates fourth-order convergence at the
low ($\rho=2$) resolution.

We conclude from the gauge wave and Gowdy wave results
that our BSSN code is stable, accurate, and convergent.

\section{Head-on Binary Black-Hole Collisions}
\label{sec:headon}

In this section we present results for head-on collisions of two
equal-mass Misner-Wheeler-Brill-Lindquist (MWBL) black
holes~\cite{Brill63,Misner57}. These results extend our previous tests
of \LAZEV\ to more interesting nonlinear spacetimes containing binary
black holes.

The MWBL data represent conformally flat slices of multiple black hole
space-times with $n$ punctures. The data are parametrized by $n$
puncture masses $m_i$ and $n$ puncture positions ${\vec r_i}$ (in the
conformal space). The ADM mass is given by the sum of $m_i$, and the
ADM linear momentum and angular momentum are zero.  The data have the
form
\begin{eqnarray}
  \gamma_{ij} &=& \psi^4 \delta_{ij}, \\
  K_{ij} &=& 0,
\end{eqnarray}
where 
\begin{equation}
  \psi = 1 + \sum_{i=1}^n \frac{m_i}{2 r_i},
\end{equation}
and $r_i = |{\vec r} - {\vec r_i}|$.

\subsection{Setup}
We used MWBL octant-symmetric data consisting of two equal mass black
holes aligned along the z-axis, allowing us to evolve the data using
the $\{x>0,y>0,z>0\}$ octant. In addition, we use a `Transition
Fisheye' transformation~\cite{Baker:2001sf} to
enlarge the physical domain without sacrificing resolution near the
punctures.  The `Transition Fisheye' transformation is a smooth radial
transformation from an inner resolution fixed by the Cactus gridspacing
($h$)
to an outer resolution that is generally lower than the inner
resolution. This transformation is parametrized by an outer
`de-resolution parameter' $a$ (the effective grid-spacing at the edge of
the grid is $a\,h$), a transition width parameter `$s$', and
the center of the transition region `$r_0$'.  The transformation has the
form
\begin{equation}
  r_{physical} = r (a + (1-a) {\cal R}(r)),
\end{equation}
where $r_{physical}$ is the physical radius corresponding to the coordinate
radius $r$ and ${\cal R}(r)$ is given by
\begin{equation}
  {\cal R}(r) = \frac{s}{2 r \tanh\frac{r_0}{s}} 
\ln\left(\frac{\cosh\frac{r+r_0}{s}}{\cosh\frac{r-r_0}{s}}\right).
\end{equation}

For these runs we set the mass parameter of the two holes to $0.5M$, and
the puncture positions to $(0,0,-1.1515M)$ and $(0,0,1.1515M)$. For most of the
runs, the computational grid extended to  $12.3M$, which corresponds to a physical
size of $26M$. We also performed some short runs where the computational grid 
extended to $24.6M$, which corresponds to a physical size of $63M$. These latter runs
were used to determine the effect of the boundaries on the waveforms, constraint violations,
and horizon mass.
The Fisheye parameters used in these runs were
$a = 3$, $ s = 1.2M$, $r_0 = 5.5M$. We performed runs at resolutions of
$h= 1.1515M / (9 \rho)$ with gridsizes of $(96 \rho)^3$
gridpoints along one octant, where $\rho=1,2,4$. The gridsizes were 
chosen so that the
punctures lie halfway between gridpoints. For the runs with the boundary at
$63M$ we used the same resolutions
but chose gridsizes of $(192 \rho)^3$ gridpoint along one octant, where
$\rho =1, 2$.

We evolved the data with  the standard {\tt 1+log} lapse, where the
initial value of $A$ (see Eq.~\ref{eq:kdriver_A}) is zero,
and a modified {\tt 1+log} lapse, where the initial value of $A$ is
\begin{eqnarray}
  A(t=0) &=& c\,\exp\left(-(\psi -1)^{-2} / \sigma\right). \label{eq:A0}
\end{eqnarray}
We choose 1 for the initial value of the lapse.
When using the modified lapse we set $c=0.5$, and $\sigma  =
0.8$.
This modified lapse, unlike the original {\tt 1+log} lapse, collapses at
the puncture in the continuum limit. Analytically the standard {\tt 1+log}
lapse should retain its initial value at the puncture throughout the
entire evolution. Since we start with a lapse of one, the lapse should
not collapse at the puncture. However, the lapse will collapse {\em
near} the
puncture, and a lack of resolution drives the lapse to
zero at the puncture as well. However, as the grid is refined this
artificial collapse at the puncture is delayed, leading to short wavelength
features and blow-ups near the
puncture. The initial value of $A$ in Eq.~(\ref{eq:A0}) forces the
lapse to collapse near the puncture in the continuum limit.
Unfortunately this modification introduces its own problems as
described below (see Fig.~\ref{fig:lapse_compare}).

We used the second form of the `Gamma-Driver' shift
Eq.~(\ref{eq:gamma_driver_alt}) to evolve the shift. The initial value of
the shift and its time derivative were zero. We used both a constant
$\eta$ ($ \eta = 2.8/M$) and
a spatially varying $\eta$ equal to $2.8/M$ at infinity and $5.6/M$ at the 
punctures, as suggested by Diener~\cite{PD_unpub}. 
The spatially varying $\eta$ was of the form~\cite{Alcubierre:2004bm}
\begin{equation}
  \label{eq:eta_var}
  \eta = \eta_p - \frac{\eta_p - \eta_{\infty}}{(\psi - 1)^2 +1},
\end{equation}
where $\eta_p$ and $\eta_{\infty}$ are parameters specifying $\eta$ at the
puncture and infinity respectively, and $\psi$ is the puncture data
conformal factor. The function $F$ in the 
`Gamma-Driver' shift was set
to $F=\frac{3}{4} \alpha / \psi^4$.

Pure fourth-order runs proved to be
very unstable near the punctures both with and without upwinded
stencils, and Kreiss-Oliger dissipation further destabilized these runs
(near the punctures).  We had success in
stabilizing these
runs with two techniques: (i) using second-order upwinding of
advection terms over the entire grid, and (ii) a localized
(within the apparent horizon) order reduction (LOR) to all second-order
accurate
spatial differentiation (including second-order upwinding of advection
terms). This
latter method provides significantly more accurate waveforms. When using
the LOR method we found that
lower-order accuracy is needed both at the punctures and at the
origin. However, we only reduce the order of accuracy near the origin
after a common apparent horizon forms (typically $5M$ after the common apparent horizon
forms). We found that Kreiss-Oliger dissipation can be used to remove
late time instabilities if the dissipation coefficient is set to zero
in the LOR region. However, we did not use dissipation in the runs
presented below.

We implement LOR in the following
way. First we calculate the time derivatives of all variables using
the standard fourth-order stencils in
Eq's~(\ref{eq:dcx})~-~(\ref{eq:dux}). Then we overwrite
all points within some given coordinate-ellipsoid with the time derivatives
calculated using the standard second-order stencils (with second-order
upwinded advection terms). The dissipation operator inside the LOR
region can be either the same one chosen for the rest of the grid or a
lower-order operator (in either case the dissipation coefficient inside
the LOR region can be different from the dissipation coefficient used
for the rest of the grid). Only the spatial finite differencing is changed;
the time integrator is the same for all gridpoints.  The size of the
ellipsoid is chosen at runtime and up to eleven ellipsoids may be
used. The user is free to choose when each ellipsoid is activated. For
example, in the head-on binary black hole collision case, 
 we specify a small ellipsoid for
each of the individual apparent horizons and a larger ellipsoid for
the common apparent horizon, where the larger ellipsoid is activated only after
the common apparent horizons forms.  The sizes of these ellipsoids can be
determined by evolving with second-order accuracy over the entire grid
and finding the sizes of the apparent horizons versus time.

For the
head-on-collision runs the LOR regions consisted of spheres of radius
$0.512M$ centered on the punctures activated at $t=2M$, as well as an
ellipsoid with semi-axes $\{0.512M, 0.512M, 1.66M\}$ activated at
$t=11.6M$ (the punctures were located on the z-axis). For comparison, the
common apparent horizon has a minimum radius of $1.56M$ and a maximum
radius of $1.88M$ at $t=11.4M$. The individual apparent horizons
are approximately spherical with radii $0.61M$ at $t=1.98M$.

We used radiative boundary conditions for all variables except $\beta^i$
(which were evolved via $\partial_t \beta^i = B^i$ on the boundary).
Table~\ref{table:rad} gives the radiative boundary condition parameters
for all variables. In Table~\ref{table:rad} `$a$' is the Fisheye
de-resolution parameter. We enforce the algebraic constraints
Eq's~(\ref{eq:detconst}),(\ref{eq:trconst}) prior to applying the boundary
conditions. Hence, the boundary points do not satisfy these constraints
exactly.

\begin{table}
\caption{Radiative Boundary Condition Parameters}
\begin{ruledtabular}
\begin{tabular}{lll}\label{table:rad}
Variables & Asymptote & speed\\ 
 \hline
$\tilde g_{ii}$ & $1$ & $1/a$ \\
$\tilde g_{ij}(i\neq j)$ & $0$ & $1/a$ \\
$\tilde A_{ij}$ & $0$ & $1/a$ \\
$\tilde \Gamma^i$ & $0$ & $1/a$ \\
$\phi$ & $(1/2) \ln(a)$ & $\sqrt{2}/a$ \\
$K$ & 0 & $\sqrt{2}/a$ \\
$\alpha$ & 0 & $\sqrt{2}/a$ \\
$B^i$ & 0 & $1/a$ \\
$A$ & 0 & $1/a$ \\
\end{tabular}
\end{ruledtabular}
\end{table}

We updated the Zorro thorn of the Lazarus Toolkit~\cite{Baker:2001sf}
 to compute the Weyl
scalars to fourth-order accuracy and to make it compatible with octant,
bitant, and quadrant symmetries in any direction as well as $\pi$
rotation symmetry about the z-axis. In addition we added a spherical
harmonic decomposition routine to generate the $\ell=2$ and $\ell=4$ 
modes shown below.

\subsection{Second-Order Accurate Results}

We first confirmed that our code can reproduce the second-order accurate
waveforms published in~\cite{Alcubierre02a}. For these runs we used standard {\tt 1+log} lapse
(i.e.\ $A(0) = 0$) and Gamma-Driver shift with $\eta=2.8$. We ran with
gridsizes of $(96 \rho)^3$  gridpoints and grid resolutions of $h=1.1515/( 9 \rho)$ for
$\rho=1,2,4$. We calculated $\psi_4$ as well as its
$(\ell=2,m=0)$ and $(\ell=4,m=0)$ components using 
Zorro~\cite{Baker:2001sf}. 
Note that the $(\ell=2,m=0)$ and $(\ell=4,m=0)$ modes of $\psi_4$ are
purely real for this test.

\begin{figure}
\begin{center}
\includegraphics[width=3.2in]{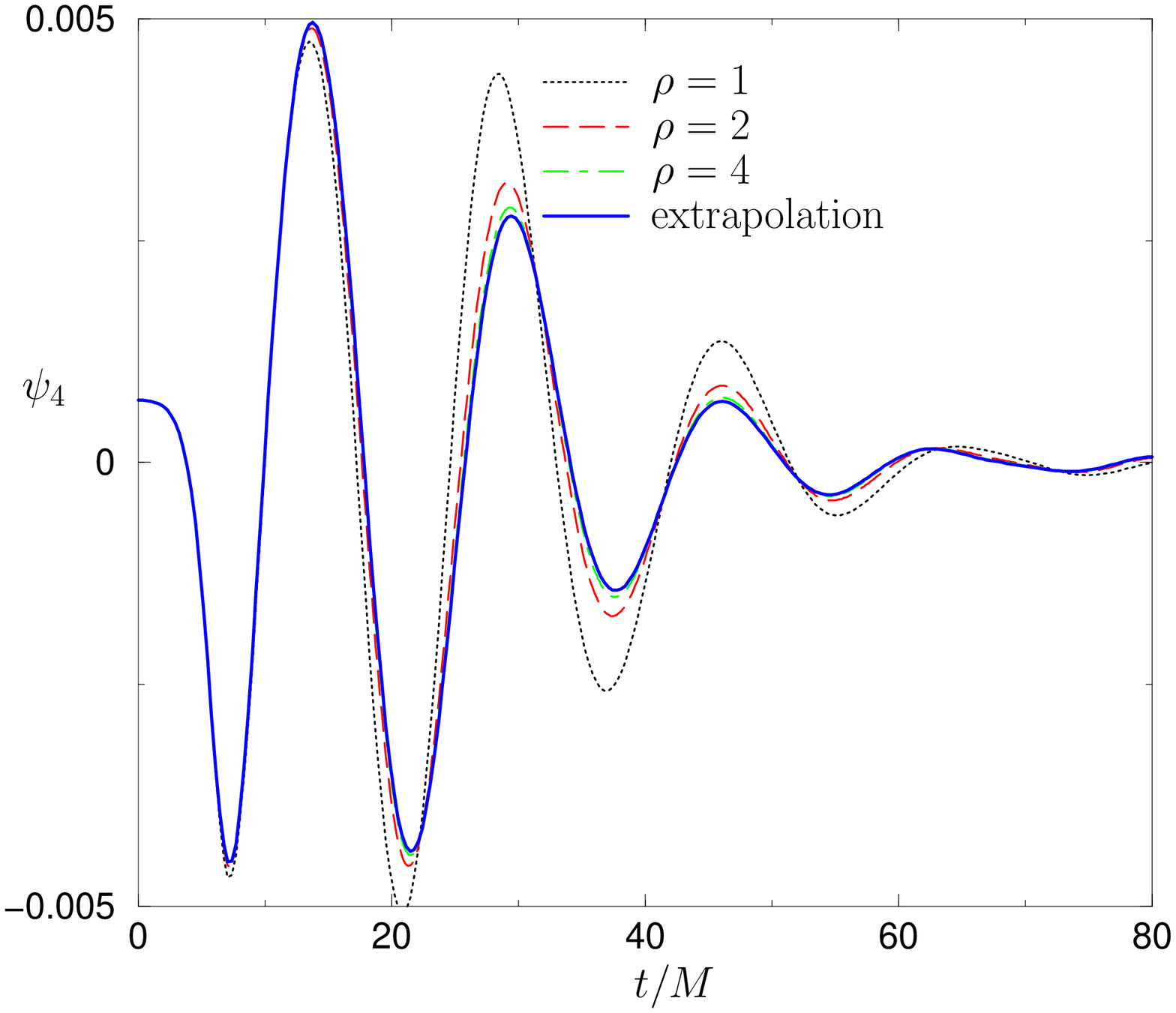}
\includegraphics[width=3.2in]{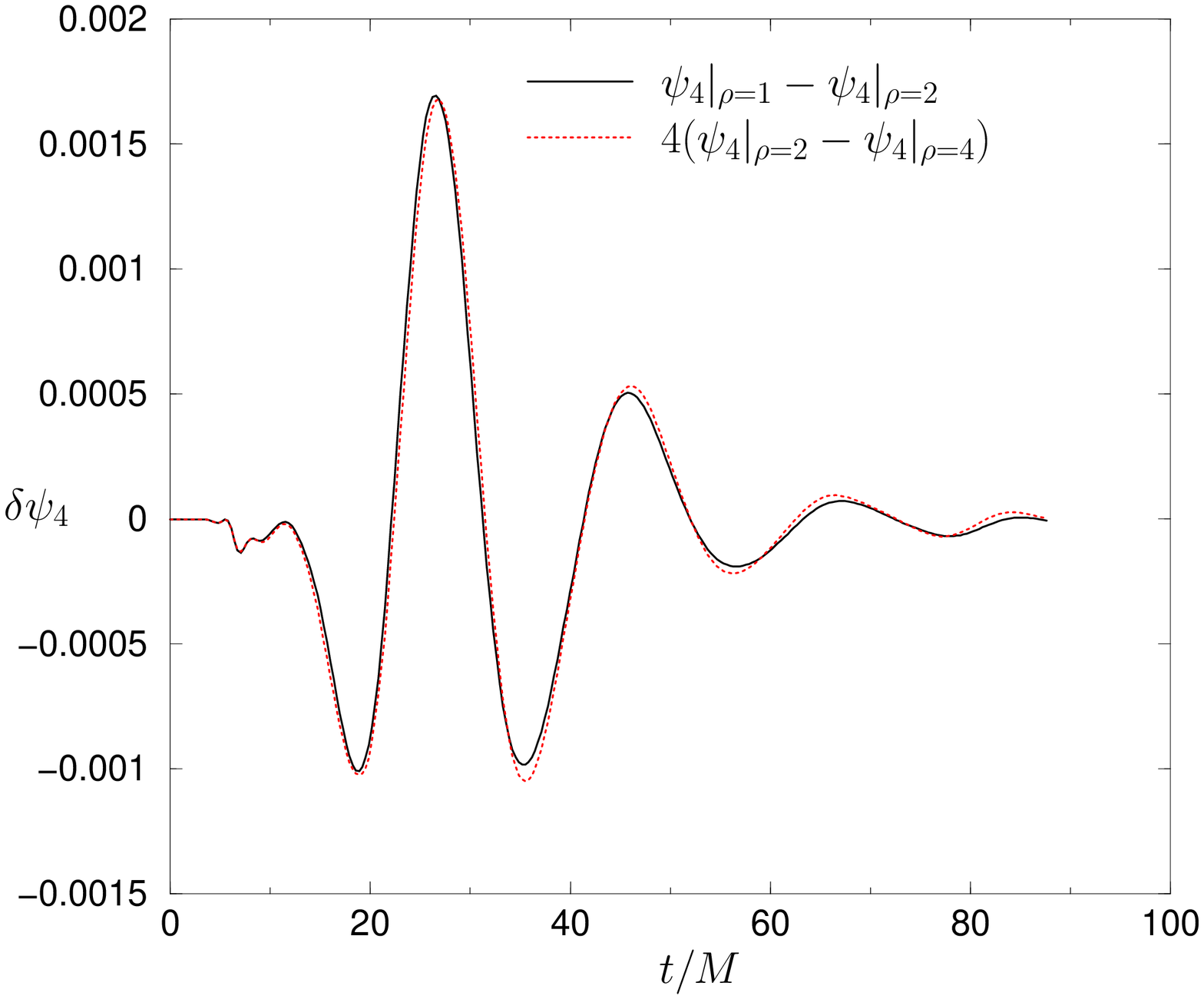}
\caption{The top plot shows the $(\ell=2,m=0)$ mode of $\psi_4$ at $r=5M$ for the
second-order evolution of the MWBL data for gridsizes $h=1.1515/(9
\rho)$, as well as the Richardson extrapolated value. The bottom plot 
shows the differences $(\psi_4|_{\rho=1} - \psi_4|_{\rho=2})$ and 
$4(\psi_4|_{\rho=2} - \psi_4|_{\rho=4})$ between these waveforms
for this mode. Note that the latter difference has been rescaled by a
factor of four. Second-order convergence is demonstrated by the
reasonable overlap between these two differences. }
\label{fig:2ndconv}
\end{center}
\end{figure}

\begin{figure}
\begin{center}
\includegraphics[width=3.2in]{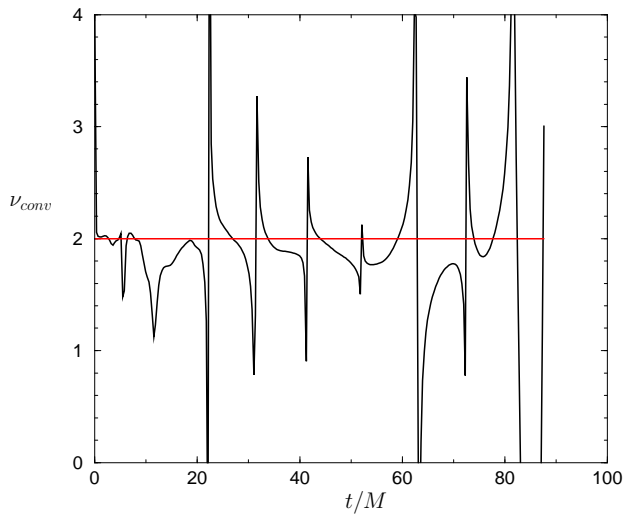}
\caption{Convergence rate of the $(\ell=2,m=0)$ mode of $\psi_4$ at
$r=5M$ for a purely
second-order evolution of MWBL head-on collision data.}
\label{fig:2ndconv_rate}
\end{center}
\end{figure}

Figure~\ref{fig:2ndconv} shows the $(\ell=2,m=0)$ mode of $\psi_4$ at
$r=5M$ for these three resolutions along with the Richardson
extrapolation of these data. 
In addition Fig.~\ref{fig:2ndconv} shows the differences
$4(\psi_4|_{\rho=2} - \psi_4|_{\rho=4})$ and 
$\psi_4|_{\rho=1} - \psi_4|_{\rho=2}$. These two difference curves overlap
  reasonably well indicating that the waveforms are
second-order accurate. However, the phase drift between the two curves
makes an evaluation of the exact order of convergence difficult.
Figure~\ref{fig:2ndconv_rate} shows the
convergence rate for this mode given by 
$$
  \nu_{conv} = \log_2 \frac{\psi_4|_{\rho=1} -
\psi_4|_{\rho=2}}{\psi_4|_{\rho=2}
- \psi_4|_{\rho=4}}.
$$
The measured convergence rate oscillates wildly but averages to about 2.

Strictly speaking $r=5M$ is not in the radiation zone so $\psi_4$ given
above does not represent the  asymptotic waveform. However, the code's
performance in calculating
the $(\ell=2,m=0)$ mode of $\psi_4$ at $r=5M$ is indicative of its
performance in calculating
this mode at large $r$.  Additionally, boundary errors
contaminate the $\ell=4$ modes at large $r$ (see Sec.~\ref{sec:4thRes}).
 Thus by extracting at
$r=5M$ we can maximize the amount of non-trivial, physically correct
quasinormal oscillations in the waveforms.


\begin{figure}
\begin{center}
\includegraphics[width=3.2in]{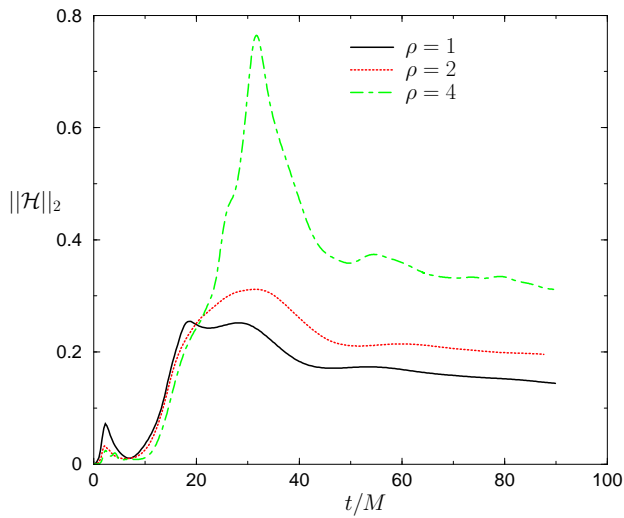}
 \caption{The $L_2$ norm of the Hamiltonian constraint violation versus
 time for the second-order
accurate head-on collision runs.  Note
that the constraint violation {\it increases } with resolution.}
\label{fig:2ndconst}
\end{center}
\end{figure}

\begin{figure}
\begin{center}
\includegraphics[width=3.2in]{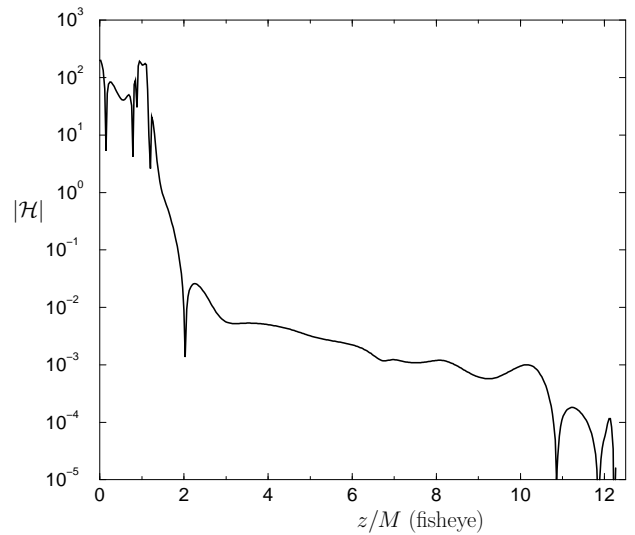}
 \caption{Hamiltonian constraint violation (absolute value) versus $z$ along the z-axis at time
$t=80M$ for the $\rho=4$ second-order head-on collision run. Note that
the large constraint violation inside the apparent
horizon ($r=2.8M$) has not leaked out into the exterior spacetime.}
\label{fig:2ndconstslice}
\end{center}
\end{figure}

Figure~\ref{fig:2ndconst} shows the $L_2$ norm of the Hamiltonian
constraint for these three runs. Note that the constraints have not been
rescaled. From the figure we can see that the constraints tend to get
bigger as the resolution is increased. The source of these constraint
violations is located between the puncture and origin. High frequency features
develop there, leading to strong constraint violations. However, as seen
in Fig.~\ref{fig:2ndconstslice} these
large constraint violations do not leak out of the apparent horizon. Thus,
although there are significant problems inside the horizon, the
region outside the horizon remains uncontaminated. 

Our calculations of the 
gravitational waveforms from the Newman-Penrose scalar $\psi_4$, 
which contains higher-order spatial derivatives of the metric, 
do not reveal any additional degree of noise and distortion with 
respect the waveforms obtained from the Zerilli-Moncrief 
formalism~\cite{Alcubierre:2004bm,Sperhake:2005uf}.

\subsection{Fourth-Order Accurate Results}
\label{sec:4thRes}
\begin{figure}
\begin{center}

\includegraphics[width=3.2in]{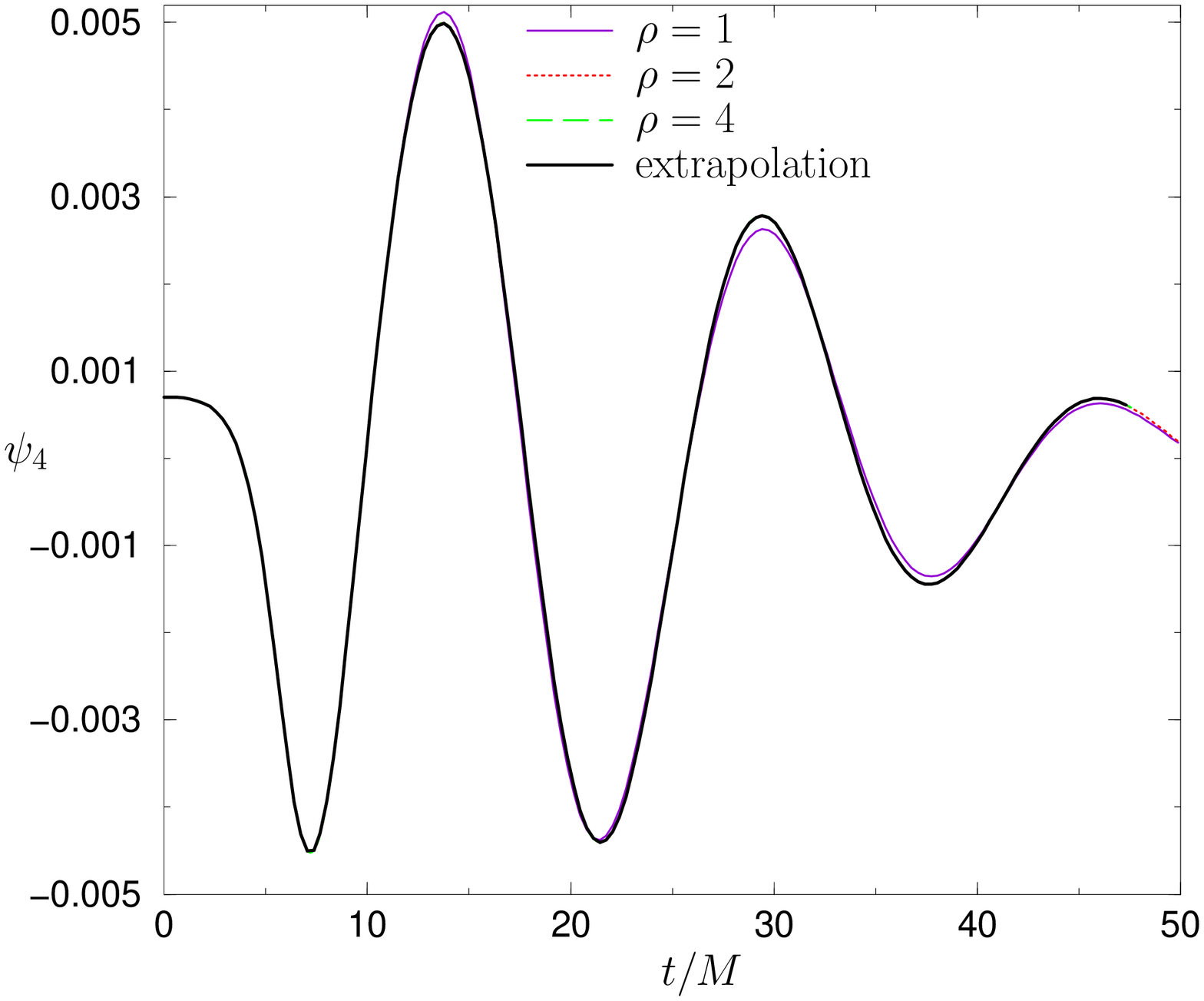}
\includegraphics[width=3.2in]{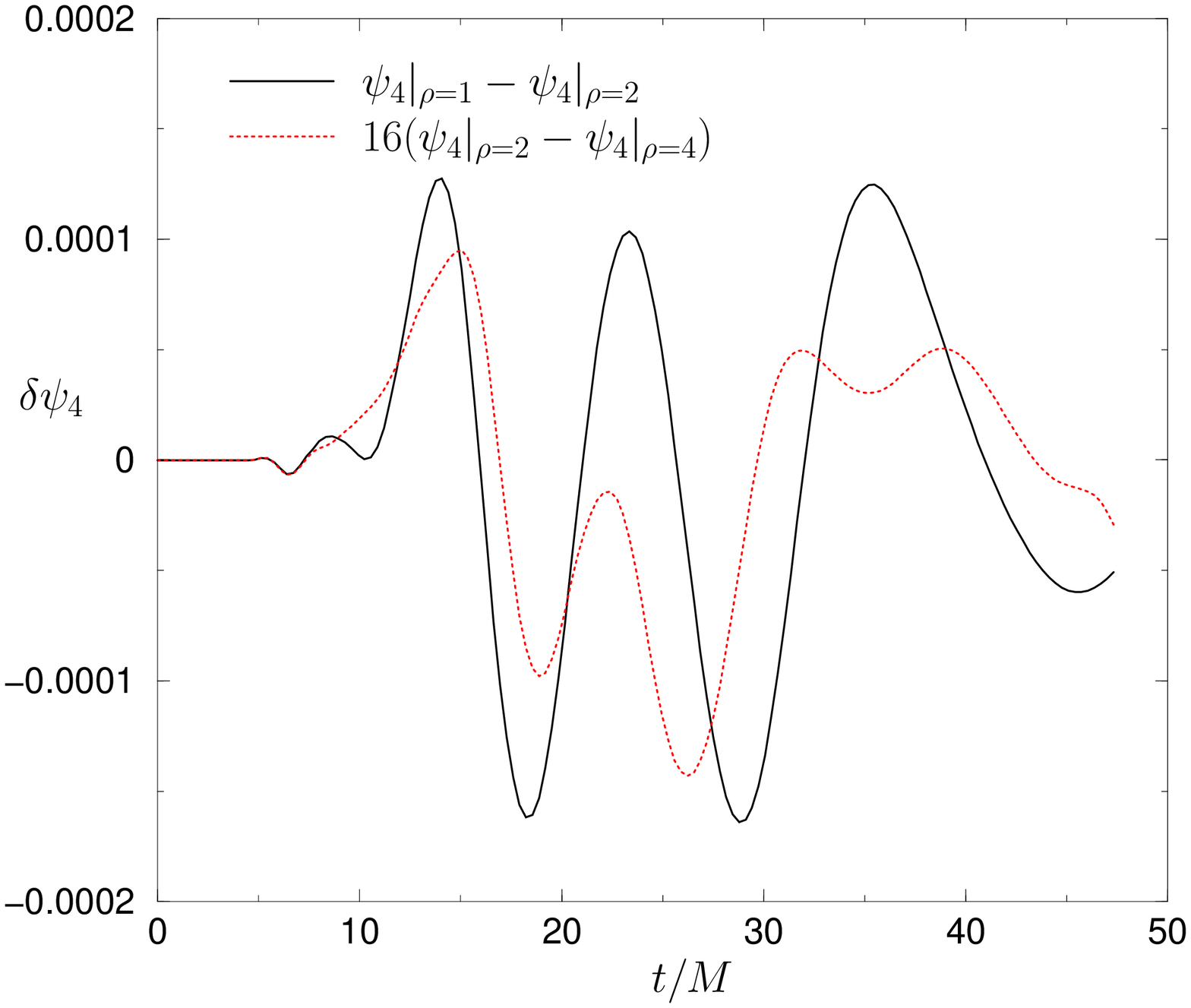}
 \caption{The top plot shows the $(\ell=2,m=0)$ mode of $\psi_4$ at
$r=5M$, produced using fourth-order evolution with LOR,
for 3 resolutions, along 
with the Richardson extrapolation curve of these data. The $\rho=2$ and
$\rho=4$ curves are indistinguishable from the extrapolated curve.
 The bottom plot shows rescaled differences 
$\psi_4|_{\rho=1} - \psi_4|_{\rho=2}$ and $16(\psi_4|_{\rho=2} -
\psi_4|_{\rho=4})$ (note the factor of 16).
The convergence rate for the fourth-order runs using LOR inside the
apparent horizons 
is not strictly fourth-order due to the phase discrepancy
between the two curves, but average waveform differences
between resolutions is consistent  with fourth-order accuracy.
Note that the amplitude of the dotted curve is less than the amplitude
of the solid curve indicating better than fourth-order reduction in the
amplitude of the errors.}
\label{fig:headon_4_conv}
\end{center}
\end{figure}

\begin{figure}
\begin{center}

\includegraphics[width=3.2in]{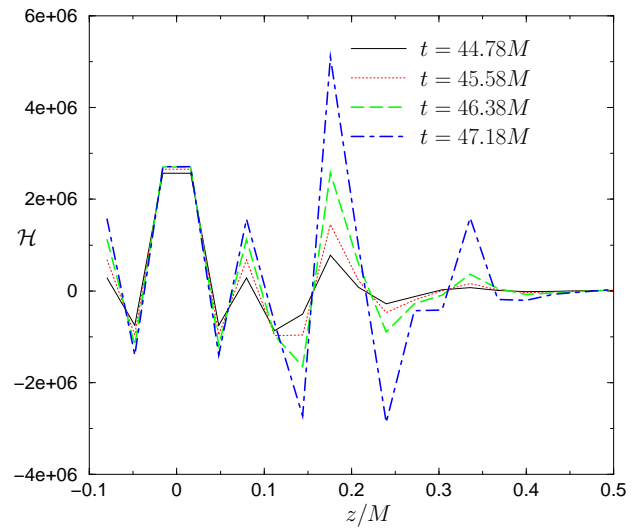}
 \caption{ The unstable mode in ${\cal H}$ for the $\rho=4$ fourth-order
runs.}
\label{fig:unstable}
\end{center}
\end{figure}

Purely fourth-order runs of puncture data proved to be unstable with
more than one black hole. We found that we could stabilize the evolution
by reducing the spatial discretization to second-order inside the apparent
horizons, and 
we successfully ran the MWBL head-on collision data with fourth-order
discretization,  fourth-order Runge Kutta time integration, and
second-order LOR regions inside the apparent horizons. 
For these runs we used the
spatially varying $\eta$ form of the Gamma-driver shift 
given by
Eq.~(\ref{eq:gamma_driver_alt}) and Eq.~(\ref{eq:eta_var}).
 Figure~\ref{fig:headon_4_conv}
demonstrates the convergence of the $(\ell=2,m=0)$ component of $\psi_4$
at $r=5M$.  Note that the system is not strictly fourth-order accurate.
There appears to be an additional phase discrepancy between the two
differences. However, the amplitude of the differences appears to be
falling at a rate consistent with better than fourth-order accuracy.
The $\rho=4$ run crashed at $47.5M$ due to an instability near the
origin (see Fig.~\ref{fig:unstable}).
This unstable mode was triggered by advection terms (the collapse of the
lapse near the origin ensures that these are the only terms which can
lead to a blowup). The fields which blew up most strongly were the 
$\tilde \Gamma^i$. This blow-up does not appear to be related to the
quadratic blow-up in $\phi$
at the puncture discussed later.  That term led to a blow up
proportional to $e^{t^2}$ (in ${\cal H}$)
near the puncture, which masked the unstable behavior near the origin. 
Both blow-ups can be modified by various choices in the gauge conditions
and it is likely that a better choice of gauge will lead to longer
fourth-order evolutions.

A fit of the $(\ell=2,m=0)$ mode of $\psi_4$ to the quasinormal
form $f\sim e^{-a t} \sin(\omega t)$ gave an exponential damping factor
of $a=0.084/M$ and frequency of $\omega =0.373/M$. The exponential damping
factor agrees to within 6\% of the expected value of
$0.0889625/M$, and
the frequency agrees to within $0.2\%$ of the expected value of
$0.373672/M$~\cite{Nollert92}.

\begin{figure}
\begin{center}

\includegraphics[width=3.2in]{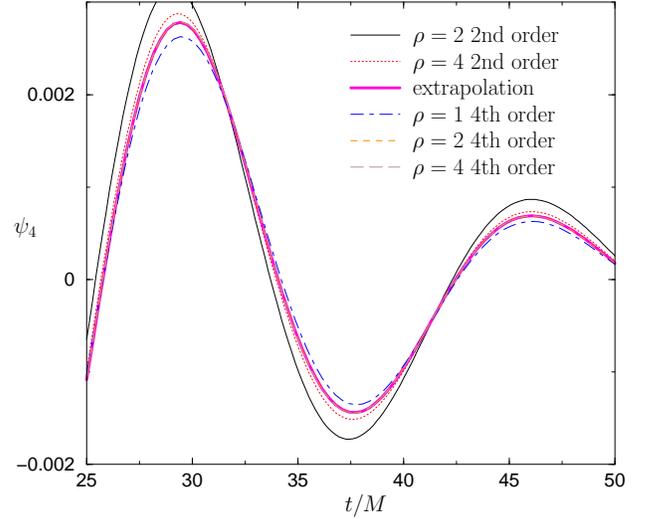}
 \caption{A comparison of the $(\ell=2,m=0)$ mode of $\psi_4$ at $r=5M$
for second and fourth-order evolutions with LOR. Note that the $\rho=1$
fourth-order waveform is better than the $\rho=2$ second-order waveform.
Similarly the $\rho=2$ fourth-order waveform is better than the
$\rho=4$ second-order waveforms. The Richardson extrapolated waveform
(based on the second-order waveforms) is
indistinguishable from the $\rho=2$ and $\rho=4$ fourth-order waveforms.}
\label{fig:headon_4_2_compare}
\end{center}
\end{figure}

\begin{figure}
\begin{center}

\includegraphics[width=3.2in]{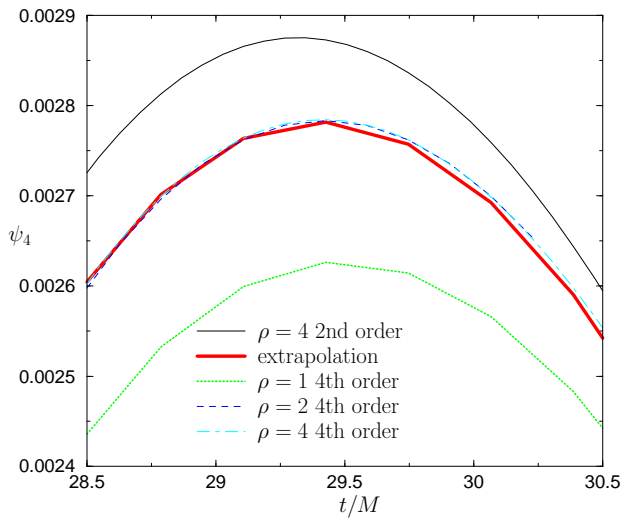}
 \caption{A comparison of the $(\ell=2,m=0)$ mode of $\psi_4$ at $r=5M$
for second and fourth-order evolutions with LOR. Note that  the $\rho=2$ 
fourth-order waveform is better than the
$\rho=4$ second-order waveforms. 
 The second-order $\rho=4$ curve lies above the
extrapolated curve while the fourth-order $\rho=1$ curve lies below.}
\label{fig:headon_4_2_compare_mag}
\end{center}
\end{figure}

In Fig.~\ref{fig:headon_4_2_compare} we show the Richardson extrapolated
value of the $(\ell=2,m=0)$ mode of $\psi_4$, calculated using the second-order accurate
results, along with the values obtained from second and fourth-order
evolutions with LOR. Note that the medium resolution fourth-order run outperforms
the high resolution second-order run. Figure~\ref{fig:headon_4_2_compare_mag} 
shows a magnified view of Fig.~\ref{fig:headon_4_2_compare} (note that
the legends in the two figures are different). 

\begin{figure}
\begin{center}

\includegraphics[width=3.2in]{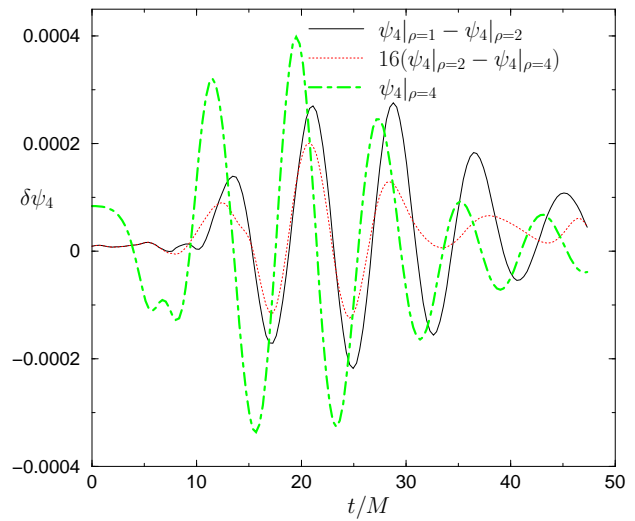}
 \caption{
The $(\ell=4,m=0)$ mode of $\psi_4$ for $\rho=4$ as well as the
differences between the $\rho=1$ and $\rho=2$ waveforms and the
$\rho=2$ and $\rho=4$ waveforms (the latter rescaled by 16).
The convergence rate of the $(\ell=4,m=0)$ mode of $\psi_4$ at
$r=5M$ for the fourth-order runs with LOR
is better than fourth-order. Note
that the error in the $\rho=1$ waveform (as evident by the difference
between the $\rho=1$ and $\rho=2$ waveforms) is approximately $50\%$ of
the amplitude of the waveform at $20M$ and larger than the waveform
beyond $30M$.}
\label{fig:l4conv}
\end{center}
\end{figure}

Figure~\ref{fig:l4conv} shows the convergence of the $(\ell=4,m=0)$
mode of $\psi_4$ at $r=5M$ calculated using fourth-order accuracy with
LOR. The plot shows the differences
$\psi_4|_{\rho=1} - \psi_4|_{\rho=2}$ and $16(\psi_4|_{\rho=2} -
\psi_4|_{\rho=4})$, as well the mode itself for $\rho=4$. Note that the
latter rescaled difference is smaller than the former. This indicates
that the $(\ell=4,m=0)$ component of the waveform is converging faster 
than fourth-order. In addition, note that the
error in the $\rho=1$ waveform is comparable to the amplitude of the
waveform. 

\begin{figure}
\begin{center}
\includegraphics[width=3.2in]{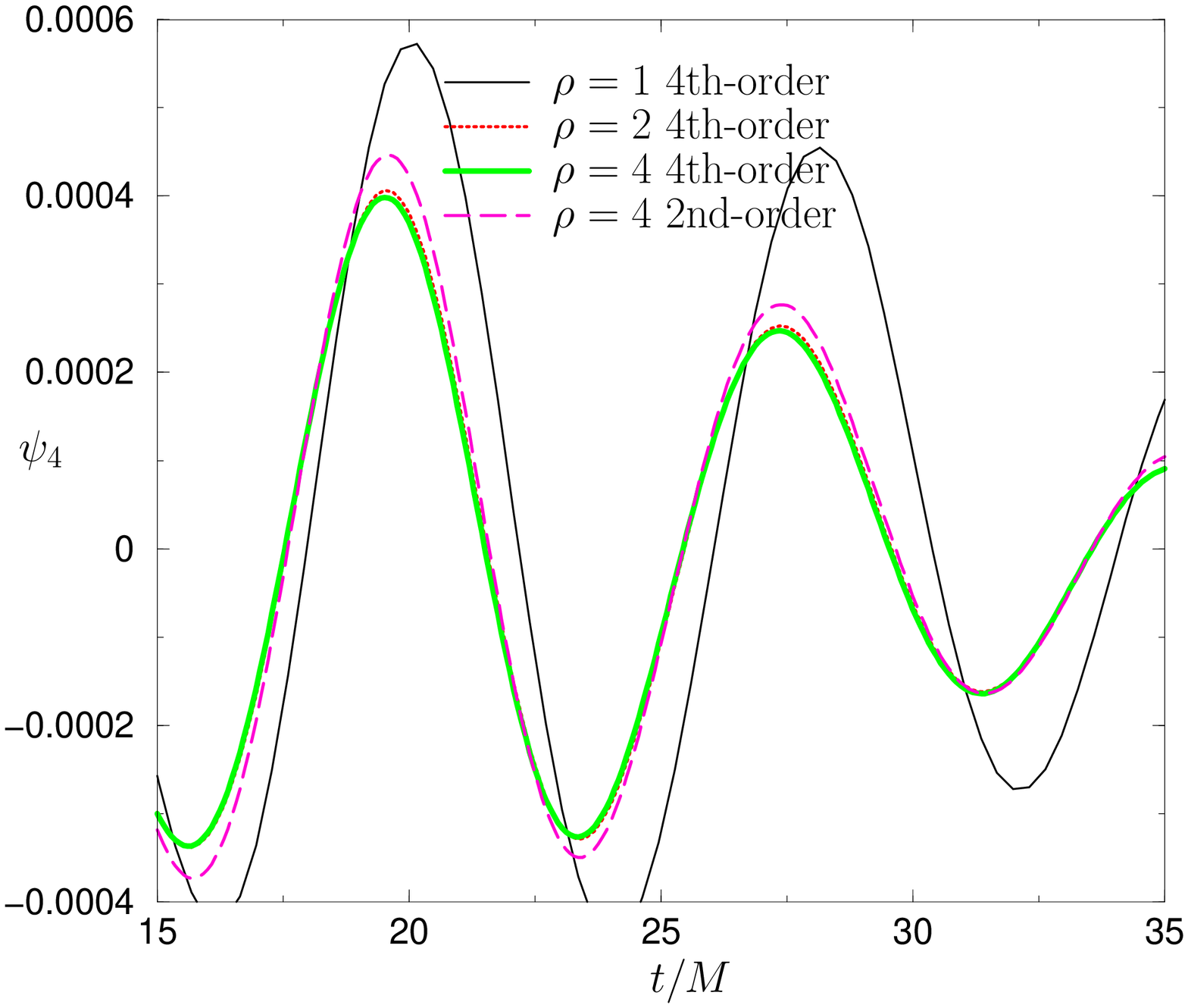}
 \caption{The $(\ell=4,m=0)$ mode of $\psi_4$ at $r=5M$ for fourth-order
evolutions with LOR as well as second-order evolutions. Note that the
$\rho=1$ fourth-order waveform has an order 100\% error and that the
$\rho=2$ fourth-order waveform has a substantially smaller error than
the $\rho=4$ second-order waveform.}
\label{fig:l4comp}
\end{center}
\end{figure}

Figure~\ref{fig:l4comp} shows a magnified view of the $(\ell=4,m=0)$ mode of $\psi_4$ at
$r=5M$ for the $\rho=1$, $\rho=2$, and $\rho=4$ fourth-order runs as
well as the $\rho=4$ second-order run. Note that the $\rho=4$
second-order results are again inferior to the $\rho=2$ fourth-order
(with LOR) results.

The $\ell=4$ modes, unlike the $\ell=2$ modes, contain significant
contamination from the boundary at late times. 
For example, when extracting at $r=13M$ we find that the amplitude of
the $(\ell=4,m=4)$
mode of $\psi_4$ (which is non-zero due to boundary errors) is
approximately $50\%$ that of the $(\ell=4,m=0)$ mode at $t=45M$.
This contamination is stronger when the extraction sphere is further
out.
However, the waveform extracted at $5M$ shows
significantly less contamination. The reason for this is that the
incoming error is delayed by about $8M$ while the outgoing mode is
advanced by $8M$ (for a net gain of $16M$ in reliable data). In addition 
the outgoing mode, which has a
$1/r$ falloff, is correspondingly larger at this smaller radius.
This boundary contamination problem can be mitigated by pushing the outer
boundary further out. This can be achieved within the existing \LAZEV\
framework using a stronger fisheye
transformation or adding more gridpoints.

Although the waveforms from the fourth-order runs converge as expected
the Hamiltonian constraint does not. High frequency features near the
puncture and origin lead to large Hamiltonian constraint violation. At
the points surrounding the puncture these high frequency features induce
a quadratic blow-up in $\phi$ for the $\rho=4$ run. This quadratic behavior, which is confined to the nearest
neighbor points to the puncture, leads to a blow-up
of $e^{t^2}$ in the Hamiltonian constraint due to terms proportional to
$e^{4 \phi}$. This blow-up in ${\cal H}$ is entirely
localized to the puncture and does not affect points outside.
In the following figures we demonstrate that the large constraint violations
inside the apparent horizon do not leak out and that the constraint
violations outside the apparent horizon converge to zero when boundary effects
are removed.

\begin{figure}
\begin{center}

\includegraphics[width=3.2in]{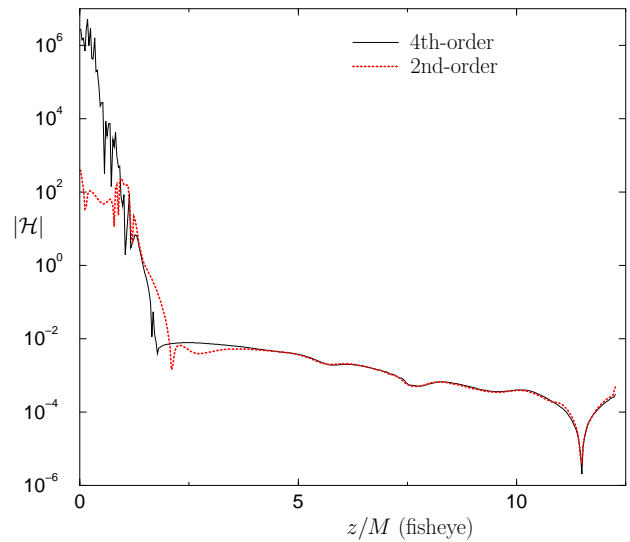}
 \caption{The absolute value of the Hamiltonian constraint along the z-axis at $t=47.2M$ for
the second and fourth-order runs. The
points near the puncture have been removed (the violation at the puncture
was ~$10^{117}$) from the fourth-order data. Note that the constraint
violation is $10^4$ times bigger inside the horizon for the 4th order
run, and that the extreme constraint violation does not leak out of the
horizon (located at $z=2.8M$).}
\label{fig:headon_4_HC_slice}
\end{center}
\end{figure}

Figure~\ref{fig:headon_4_HC_slice} shows the Hamiltonian constraint
along the z-axis (the points surrounding the puncture have been removed)
at $t=47.2M$
for the second and fourth-order runs for $\rho=4$. The Hamiltonian constraint inside
the horizon is as much as $10^4$ times larger for the fourth-order run.
However, outside the horizon the constraint violations are very similar.
Note that we do not expect the constraints to converge to zero in this
case because our boundary conditions are not constraint preserving and
boundary constraint violations have contaminated the solution at this time
(the boundaries were at $26M$ in physical coordinates).
The $\rho=4$ fourth-order run with LOR crashed due to an instability near
the origin. Figure~\ref{fig:unstable} shows the unstable mode in ${\cal
H}$.

\begin{figure}
\begin{center}

\includegraphics[width=3.2in]{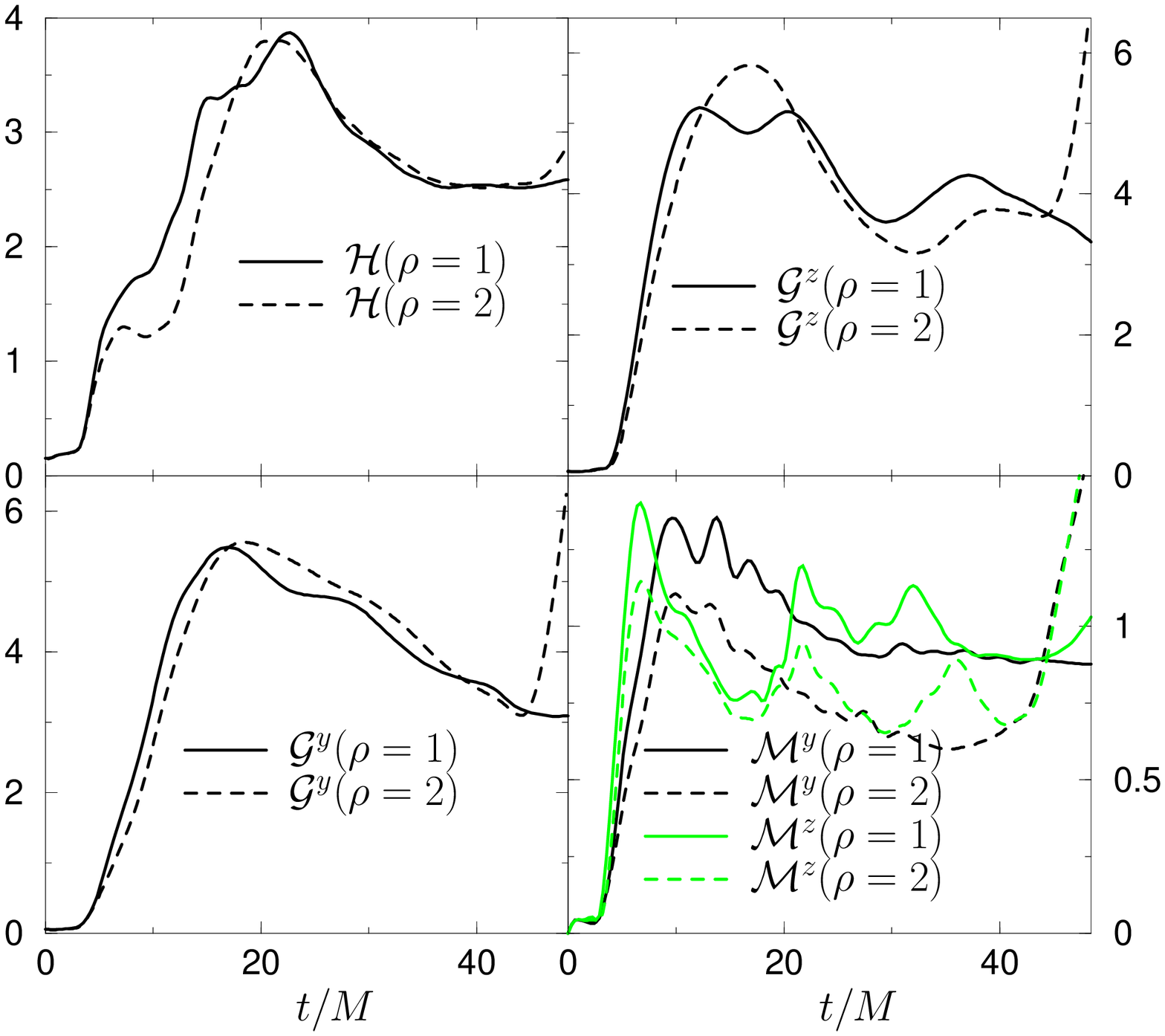}
\caption{The $L_2$ norm, restricted to the region $3M <
r_{physical} < 26M$, of all constraint violations. The boundaries
were located at $63M$. All norms have been multiplied by $10^5$ and the
$\rho=2$ norms have been rescaled by an additional factor of 16.
In each panel the solid curves correspond to
$\rho=1$ and the dashed curves to $\rho=2$. The reasonable overlap of
 the dashed and solid curves
indicate that the constraint violations are converging to zero to
fourth-order. The constraints no longer converge to zero after
$t\sim43M$ due to boundary effects.}
\label{fig:constraint_plot}
\end{center}
\end{figure}

 In
Fig.~\ref{fig:constraint_plot} we show the $L_2$ norms of the
Hamiltonian constraint ${\cal H}$, momentum constraint ${\cal M}^i$, and
the BSSN constraints ${\cal G}^i$ (note that the $x$ and $y$ components
of the both momentum and BSSN constraints are equal). We restricted
these norms to the region  outside $r_{physical}=3M$ (the horizon is at
$r_{physical}\sim 2M$) and inside
$r_{physical}=26M$, where $r_{physical}$ is the physical radius. The outer boundaries
were located at $63M$ in physical coordinates. These runs required double the number of
gridpoints (along each direction) as the standard runs for the same resolutions.
The reasonable overlap of the $\rho=1$ and rescaled $\rho=2$ curves for $43M$ of
evolution
indicates that the $L_2$ norms are approximately fourth-order convergent (the convergence
is lost near $t=40M$ due to constraint violating modes propagating in from the boundary).
In addition, the amplitude of the constraint violations is $\sim 10^{-5}$,
which is nine order of magnitude smaller than the maximum Hamiltonian
 violation inside the horizon.
We conclude from this figure that the second-order errors introduced by
the LOR differencing, as well as the extremely large Hamiltonian constraint
violation inside the apparent horizon, do not propagate outside of 
the apparent horizon.

\begin{figure}
\begin{center}
\includegraphics[width=3.2in]{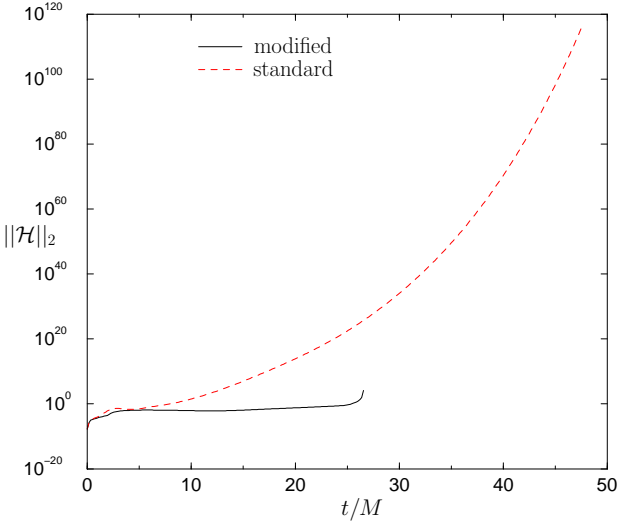}
 \caption{The $L_2$ norm of the Hamiltonian constraint violation for the
$\rho=4$ fourth-order evolution with LOR for the standard and modified {\tt
1+log} lapses. Note the $e^{t^2}$ blow-up in the run using the standard
{\tt 1+log} lapse and that the run using the modified {\tt 1+log} lapse
crashes at $26.5M$.}
\label{fig:lapse_compare}
\end{center}
\end{figure}

The extreme Hamiltonian violation near the puncture can be removed using
the modified {\tt 1+log} lapse. However, this modification tends to
destabilize the runs at late times. Figure~\ref{fig:lapse_compare} shows
the $L_2$ norm of the Hamiltonian constraint violation  for the $\rho=4$
fourth-order run with LOR. The modified lapse produces a
constraint violation smaller than 1 for $25M$ but also causes the run to
crash at $26.5M$. The standard {\tt 1+log} lapse produces a ``stable''
evolution to  $47.5M$ (see above) but has a constraint violation
proportional to $e^{t^2}$ at the puncture. If short term evolutions,
like the ones required for the Lazarus techniques, are required, then
the modified {\tt 1+log} is preferred. 

\begin{figure}
\begin{center}

\includegraphics[width=3.2in]{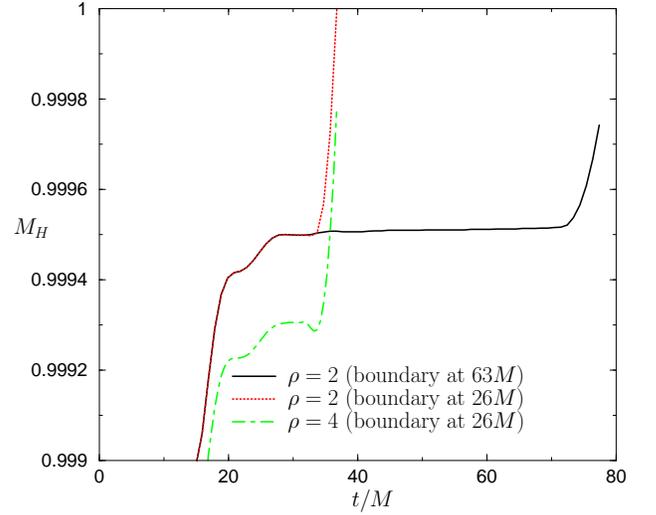}
 \caption{The horizon mass versus time for the fourth-order runs with LOR.
The thick solid line and the dotted line show the horizon mass versus time
for the $\rho=2$ resolution with the physical boundary at $63M$ and $26M$
 respectively.
Note that the sharp increase in mass at later times is due to boundary
effects. The dashed line shows the horizon mass for the $\rho=4$ run with the
boundaries at $26M$. The error in the horizon mass at $t=63.5M$ is
$2.6\%$ for the $\rho=2$ run with boundary at $26M$.}
\label{fig:mass_v_time}
\end{center}
\end{figure}

Figure~\ref{fig:mass_v_time} shows the horizon mass versus time, as
calculated by the AHFinderDirect thorn~\cite{Thornburg2003:AH-finding},
for the $\rho=2$ and $\rho=4$ runs with LOR. The plot also shows the
horizon mass of a $\rho=2$ run with double the standard number of
gridpoints in each direction and outer boundary at $63M$ in physical coordinates.
 The relatively sharp
increase in the horizon mass at $t=32M$ is due to contamination from the
boundary. When the boundary is moved out to $63M$ (from $26M$) this increase is
 delayed by roughly $\Delta t=37M$. Also note that the slope of the increase is reduced when
the boundary is further out. The measured convergence rate of the
horizon mass (based on runs with the boundary at $26M$) 
exceeded 3.8 for $30M$ of evolution.

\begin{figure}
\begin{center}
\includegraphics[width=3.2in]{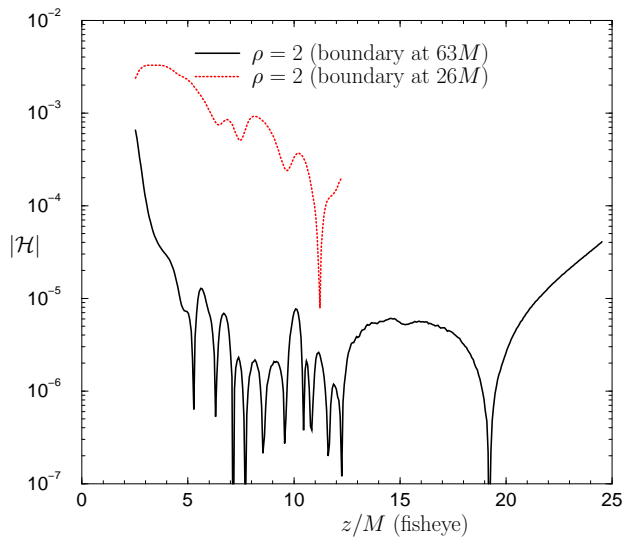}
 \caption{The Hamiltonian constraint along the $z$ axis at $t=40.8M$
for the fourth-order (with LOR) runs with gridspacing $h=1.1515/18$
(i.e.\ $\rho=2$). $z_{fisheye}=12.3M$ corresponds to $26M$ in physical
coordinates and $z_{fisheye}=24.6M$ corresponds to $63M$ in physical coordinates. 
 Boundary
contamination for the larger run has just reached $z_{fisheye}=12M$.
Note that in the range $5M < z_{fisheye} < 12 M$ the Hamiltonian
constraint is 100 time smaller when the boundary is pushed out to $63M$.}
\label{fig:HC_bdry_compare}
\end{center}
\end{figure}

The horizon-mass versus time plot shows that the late time boundary
contamination contains significant erroneous gravitational radiation. To
determine when the calculated gravitational waveforms are good
approximations to the true waveforms, we need to confirm that (i) the
waveforms converge and do not change significantly when the resolution
is further increased, (ii) the waveforms do not change significantly when 
the boundary effects are removed (which we achieved by causally 
disconnecting the  boundaries from the observer), and (iii) the
Hamiltonian constraint violation (inside the extraction region) is much
smaller than the waveform amplitude. We have shown that the waveforms
converge, and that the Hamiltonian constraint violation converges in the
extraction region (see Fig.~\ref{fig:constraint_plot}).
Here we show the effects of the boundary on the Hamiltonian constraint and
the waveforms.
Figure \ref{fig:HC_bdry_compare} shows
Hamiltonian constraint violation along the $z$ axis at time $t=40.8M$
for the standard $\rho=2$ run and a run with twice the standard
number of gridpoints.  The boundary in this latter case was at $63M$
(note that the $z$-axis plots the fisheye coordinate and that
$z_{fisheye}=12.3M$ corresponds to the physical coordinate
$z_{physical}=26M$ and $z_{fisheye}=24.6M$ corresponds to $z_{physical}=63M$). 
The figure shows the Hamiltonian constraint outside the apparent horizon. Note
that at this time the boundary errors for the larger run have just
reached $z_{fisheye}=12M$. The Hamiltonian constraint, in the range
$z_{fisheye}=5M$ to $z_{fisheye}=12M$, is 100 times
smaller for the run with the larger boundary. Even at $t=70M$,
when the boundary errors
have contaminated the entire grid, the Hamiltonian constraint is 20
times smaller for the larger run in this range.

\begin{figure}
\begin{center}
\includegraphics[width=3.2in]{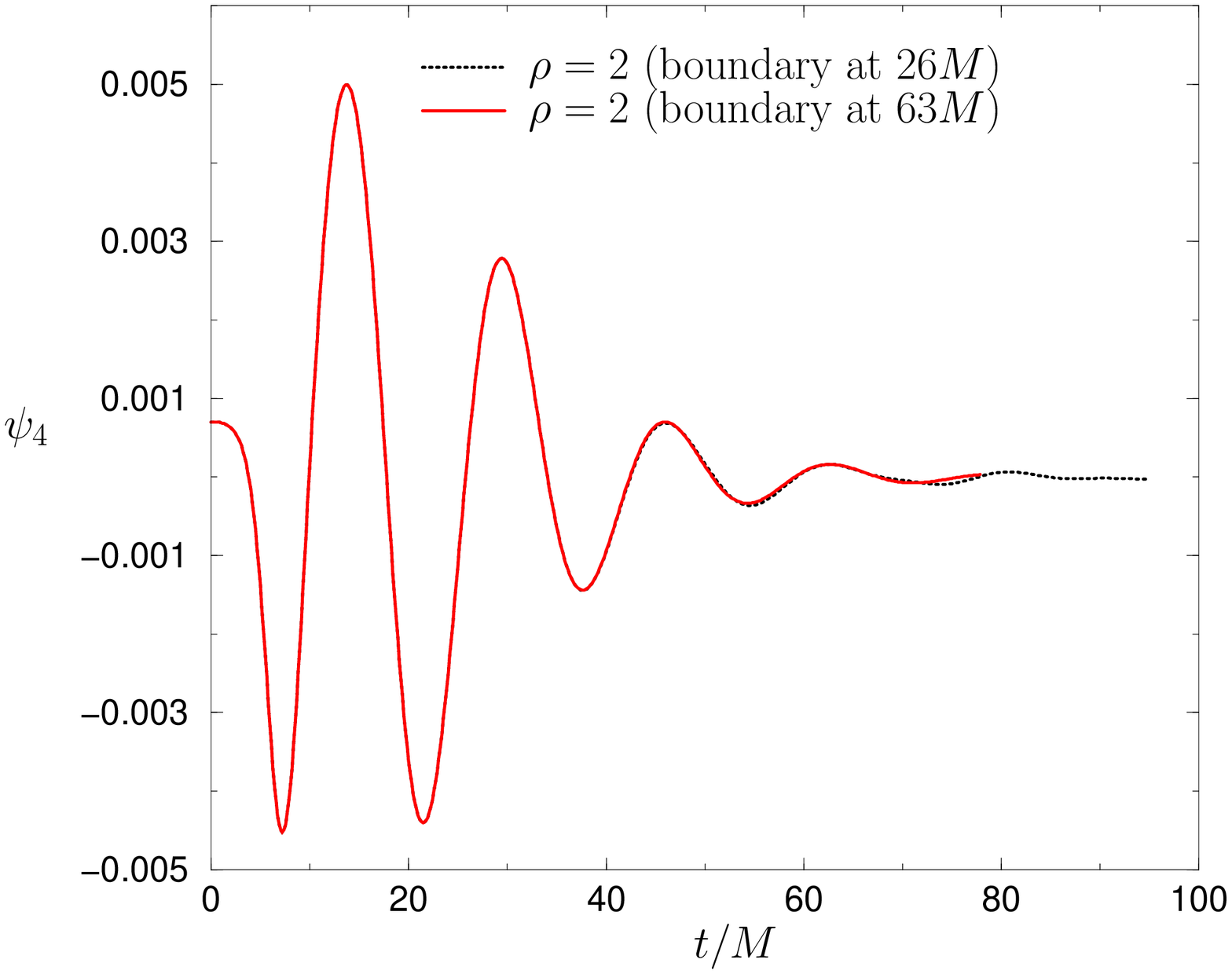}
 \caption{The $(\ell=2,m=0)$ mode of $\psi_4$ (observer at $r=5M$) for
the fourth-order (with LOR) $\rho=2$ run with the physical boundary at
$26M$
 (standard) and
$63M$. The effect of the boundary is not significant until $t=55M$. }
\label{fig:l2_bdry_comp}
\end{center}
\end{figure}

\begin{figure}
\begin{center}
\includegraphics[width=3.2in]{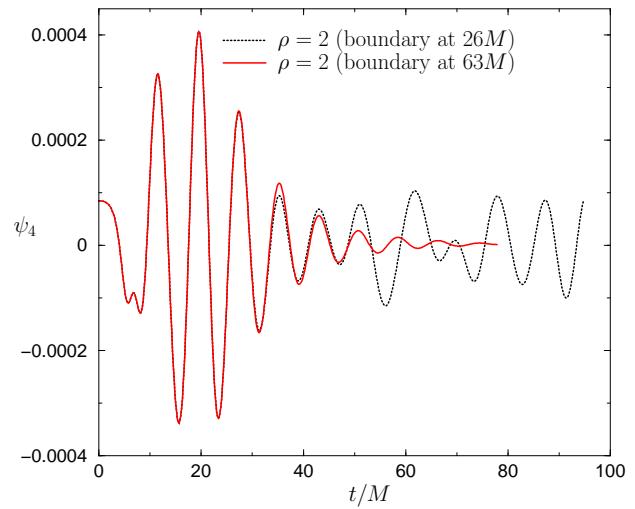}
 \caption{The $(\ell=4,m=0)$ mode of $\psi_4$ (observer at $r=5M$) for
the fourth-order (with LOR) $\rho=2$ run with the physical boundary at
$26M$ (standard) and
$63M$. The effect of the boundary is significant at $t>33M$.}
\label{fig:l4_bdry_comp}
\end{center}
\end{figure}

 As seen in
Fig.~\ref{fig:l2_bdry_comp}, the effect of reducing the boundary
noise is not readily apparent in the $(\ell=2,m=0)$ mode of $\psi_4$.
However, as seen in Fig.~\ref{fig:l4_bdry_comp}, 
the effect is readily observable in the $(\ell=4,m=0)$ mode. We
conclude, based on these two figures, that the $(\ell=2,m=0)$ mode is
represented accurately to $55M$ when the boundaries are located at $26M$,
 while the  $(\ell=4,m=0)$ mode is
represented accurately only to $33M$ for the same boundary location.
The $(\ell=2,m=0)$ mode is more accurate
because the boundary contamination contains relatively high frequencies
which are filtered more effectively by the $(\ell=2,m=0)$ angular
integration. Note that we placed the observer at $r=5M$ and that the
 waveforms become inaccurate sooner when the extraction
sphere is placed at larger radii.

\begin{figure}
\begin{center}
\includegraphics[width=3.2in]{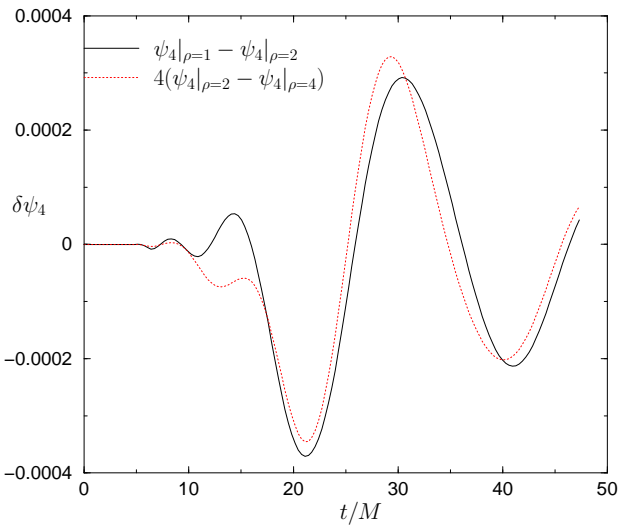}
 \caption{Convergence plot for the mixed fourth-ordered centered differencing
with second-order upwinded differencing runs. Note that the difference
$\psi_4|_{\rho=2} - \psi_4|_{\rho=4}$ has been multiplied by 4, indicating
approximate second-order convergence.}
\label{fig:42upconv}
\end{center}
\end{figure}

\begin{figure}
\begin{center}

\includegraphics[width=3.2in]{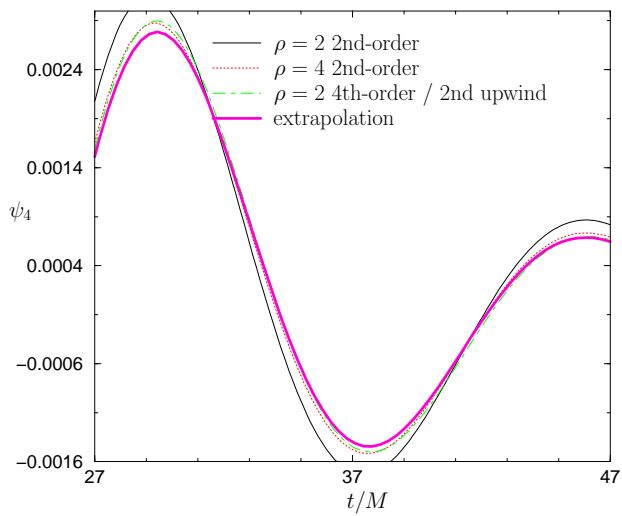}
 \caption{A comparison of the $(\ell=2,m=0)$ mode of $\psi_4$ at $r=5M$
produced using the standard second-order evolution and a mixed
fourth-order with second-order upwinding evolution. Note that the
$\rho=2$ waveform from the mixed fourth-order/second-order upwinding is
of similar quality to the $\rho=4$ second-order waveform. The
extrapolated values used in this plot are based on the waveforms from
the fourth-order runs with LOR.}
\label{fig:42upcomp}
\end{center}
\end{figure}

We conclude this section by showing the waveforms produced with
fourth-order centered spatial finite differencing for all terms except
the advection terms, for which we used second-order
upwinded differencing. The waveforms produced by this method are
inferior to those produced with LOR techniques but superior to those
produced by the standard second-order technique.
 The time integration was carried out with the
standard fourth-order Runge Kutta integrator. 
Figure~\ref{fig:42upconv} shows the differences
$\psi_4|_{\rho=1} - \psi_4|_{\rho=2}$ and $\psi_4|_{\rho=2} -
\psi_4|_{\rho=4}$ with the latter rescaled by 4 (to indicate second-order
convergence) for the $(\ell=2,m=0)$
mode of $\psi_4$ at $r=5M$. The two curves do not
overlap exactly due to phase drift. Despite the second-order accuracy of the
algorithm, the waveforms produced with this technique are more accurate
than those produced by the purely second-order spatial differencing.
Figure~\ref{fig:42upcomp} shows the $(\ell=2,m=0)$
mode of $\psi_4$ at $r=5M$ for the standard second-order evolution as
well as the mixed fourth-order with second-order upwinding evolution.
Note that the $\rho=2$ waveform from the latter technique is of a
similar quality to the $\rho=4$ waveform from the purely second-order
technique. The $\rho=4$ run crashed at $47.5M$ with the same mode that
killed the $\rho=4$ LOR run. Both techniques crashed at
the same time and in the same way because the instability near the
origin is driven
by advection terms, and both techniques use the same upwinded advection
stencil near the origin. Figure~\ref{fig:unstable} shows the unstable
mode (along the z-axis) that crashes the $\rho=4$ fourth-order runs
(with LOR and with second-order upwinding).

We computed the energy radiated from the head-on collision by several
different methods. We first estimated it by computing the difference
between the final horizon mass and the total ADM mass, and obtained a
radiated energy in the range $(7-8)\times10^{-4}M$.  We also used the
Lazarus method to extract Cauchy data for the Teukolsky equation at
relatively early evolution times $T\leq20M$, and obtained
$E_{radiated}=(8\pm1)\times10^{-4}M$. The direct integration of the
$\ell=2$ and $\ell=4$ waveforms that we present here, extracted at numerical radii between
$5M$ and $10M$, produce energy estimates of the order
$\sim6.6\times10^{-4}M$.

\section{Conclusion} \label{sec:conclusion}

We developed a new framework, \LAZEV, for evolving the Einstein
equations using 3+1 decompositions.  \LAZEV\ is capable of evolving with 
arbitrary-order finite difference stencils along with second, third, and
fourth-order time integrators. The overall \LAZEV\, design has a 
few novel features which
will improve significantly upon previous setups which have been
used in the Lazarus approach~\cite{Baker00b,Baker:2001sf,
Baker:2001nu,Baker:2002qf,Baker:2004wv}. \LAZEV\, is a flexible and
modular evolution package, well-suited for rapid
development and experimentation with well-posed hyperbolic systems of
evolution equations, new `live' gauge
conditions, and sophisticated boundary conditions using arbitrary
order finite differencing.

We implemented the BSSN formulation using this new framework and 
demonstrated that this new code
passes the Apples with Apples testsuite and that the code reproduces
 the second-order
accurate head-on binary black hole collisions waveforms published
in Ref.~\cite{Alcubierre02a}. We found that fourth-order accurate
evolutions of the same
data were not stable without some reduction of order and that the
most accurate waveforms were obtained by evolving the data with
second-order accurate stencils inside the apparent horizons and
fourth-order stencils outside the horizon. In that case we found that
the waveforms were fourth-order convergent and that the
fourth-order accurate $192^3$ gridpoint runs outperformed the
second-order-accurate $384^3$
gridpoint runs.  This means that our
fourth-order evolutions give  better quality waveforms with over an
 order-of-magnitude smaller computational
expense when compared to the second-order evolutions.

We also found that using a
second-order upwinded stencil for the advection terms was sufficient to
stabilize the runs. However, this introduces a second-order error in the
waveform. Nevertheless, the waveforms produced by this latter method are
superior to those produced by ordinary second-order accurate evolutions.

In this paper we demonstrated that the \LAZEV\ framework can be used
to evolve binary black hole spacetimes. We plan to extend this work
to include orbiting black holes starting from initial data based on
the conformal thin-sandwich formulation~\cite{hannam_cts} as well as
study recoil velocities from unequal mass head-on
collisions~\cite{Anninos98a}. We also
plan to extend the framework to include excision, fixed mesh
refinement, constraint damping~\cite{Marronetti:05,Berger:2004dd},
and constraint preserving boundary conditions.


\acknowledgments 
We thank Peter Diener for providing the Gamma driver shift parameters
needed to stabilize the fourth-order BBH runs. We thank Manuel Tiglio
and Peter Diener for helpful discussions on fourth-order evolutions.

We thank Peter Diener, Mark Hannam, and Bernard Kelly for helpful
discussions and for carefully reading this manuscript.

We gratefully acknowledge the support of the
NASA Center for Gravitational Wave Astronomy at University of Texas at
Brownsville (NAG5-13396) and the NSF for financial
support from grants PHY-0140326 and PHY-0354867. Computational resources
were provided by the Funes cluster at UTB.

\bibliographystyle{apsrev}
\bibliography{../../bibtex/references,local}
\thebibliography{article1}
\end{document}